\providecommand{\columnwiselinenumbersfalse}{} 
\providecommand{\runningpagewiselinenumbers}{} 
\providecommand{\linenumbers}{} 
\providecommand{\nolinenumbers}{}
\documentclass[longauth, desactivate]{aa}  

\usepackage{graphicx}
\usepackage{txfonts}
\usepackage{lipsum}
\usepackage{rotating}
\usepackage{subcaption}         
\usepackage{lscape}             
\usepackage{placeins}           

\usepackage{natbib}
\usepackage{comment}
\usepackage[dvipsnames]{xcolor}
\usepackage[colorlinks=true, citecolor=blue]{hyperref}

\newcommand{\achrene}[1]{\textcolor{Magenta}{\textsf{[#1]}}}

\begin{document}

   \title{JWST-SUPER I: New insights into irradiated warm Neptunes atmospheres from MIRI observations of HD~106315~c}
   \titlerunning{Atmospheric chemistry of HD~106315~c}


   \author{A. Dyrek\inst{1}
        \and R. Waters\inst{2,3,4}
        \and S.-M. Tsai\inst{5}
        \and M. Holmberg\inst{1}
        \and B. Edwards\inst{4}
        \and J. Pye\inst{6}
        \and M. G\"udel\inst{7,8,9}
        \and N. Crouzet\inst{10,11}
        \and R. van Gompel\inst{3}
        \and J. Polman\inst{12}
        \and F. Ardevol Martinez\inst{13}
        \and D. Barrado\inst{14}
        \and W. De Gruijter\inst{15}
        \and Th. Henning\inst{16}
        \and S. Kendrew\inst{17}
        \and N. Khorshid\inst{18}
        \and Y. Miguel\inst{11}
        \and M. Min\inst{4}
        \and M. Morales-Calderon\inst{15}
        \and M. Topinka\inst{19}
        }

   \institute{Space Telescope Science Institute, 3700 San Martin Drive, Baltimore, MD 21218\\
             \email{adyrek@stsci.edu}
        \and Department of Astrophysics/IMAPP, Radboud University, PO Box 9010, 6500 GL Nijmegen, The Netherlands 
        \and Radboud University PO box 9010, 6500 GL Nijmegen, The Netherlands
        \and SRON, Space Research Organisation Netherlands, Niels Bohrweg 4, 2333 CA Leiden, The Netherlands
        \and Institute of Astronomy \& Astrophysics, Academia Sinica, Taipei 10617, Taiwan
        \and School of Physics \& Astronomy, Space Park Leicester, University of Leicester, 92 Corporation Road, Leicester LE4 5SP, UK
        \and University of Vienna, Department of Astrophysics, T\"urkenschanzstrasse 17, A-1180 Vienna
        \and ETH Z\"urich, Institute for Particle Physics and Astrophysics, Wolfgang-Pauli-Str. 27, 8093 Z\"urich, Switzerland
        \and ASTRON, Netherlands Institute for Radio Astronomy, Oude Hoogeveensedijk 4, 7991 PD Dwingeloo, The Netherlands
        \and Université Paris-Saclay, Université Paris Cité, CEA, CNRS, AIM, 91191 Gif-sur-Yvette, France
        \and Leiden Observatory, Leiden University, P.O. Box 9513, 2300 RA Leiden, The Netherlands
        \and Division of Space Research and Planetary Sciences, Physics Institute, University of Bern, Gesellschaftsstrasse 6, 3012 Bern, Switzerland
        \and Department of Astrophysics, American Museum of Natural History, New York, NY 10024, USA
        \and Centro de Astrobiología (CAB), CSIC-INTA, ESAC Campus, Camino Bajo del Castillo s/n, 28692 Villanueva de la Cañada, Madrid, Spain 
        \and Anton Pannekoek Institute for Astronomy, University of Amsterdam, 1090GE Amsterdam, Netherlands
        \and Max-Planck-Institut f\"ur Astronomie, K\"onigstuhl 17, 69117 Heidelberg, Germany
        \and European Space Agency, Space Telescope Science Institute, 3700 San Martin Drive, Baltimore, MD 21218, USA
        \and Department of Space, Earth and Environment, Chalmers University of Technology, Gothenburg 412 96, Sweden
        \and Charles University, Faculty of Mathematics and Physics, Astronomical Institute, V Hole\v{s}ovi\v{c}k\'{a}ch 2, 182 00, Prague, Czech Republic
        }

    \authorrunning{Dyrek et al.}


 
  \abstract
   {Sulphur-bearing molecules have recently emerged as powerful tracers of atmospheric photochemistry in exoplanets observed with the James Webb Space Telescope (JWST). In several warm giant planets, SO$_2$ has been detected as a product of UV-driven chemical processing, suggesting a close connection between stellar irradiation, atmospheric metallicity, and sulphur chemistry. Whether these trends extend to Neptune-mass planets orbiting hotter stars remains largely unexplored.}
   {We investigate the atmospheric composition of the warm Neptune HD~106315~c, a Neptune-mass planet orbiting an F-type host star and subjected to a strong ultraviolet (UV) irradiation environment.} 
   {Here, we report the low-resolution (R$\sim$100) transmission spectrum between 5 and 12 $\mu$m of HD~106315~c obtained with the Mid-Infrared Instrument (MIRI) Low-Resolution Spectrometer on-board JWST. Our work also includes re-analysis of archival data from the Hubble Space Telescope (HST) Wide Field Camera 3 (WFC3) G141 mode of HD~106315~c, as well as contemporaneous XMM and \textit{Swift} monitoring of the star in the UV and X-ray wavelength ranges.}
   {Together with the archival HST WFC3 data, we detect H$_2$O and find tentative evidence for NH$_3$, retrieving abundances of $\log_{10}$(H$_2$O)$=-1.40^{+0.40}_{-0.74}$ and $\log_{10}$(NH$_3$)$=-2.47^{+0.57}_{-0.89}$, together with an isothermal terminator temperature of $747^{+150}_{-155}$~K. We place stringent upper limits on CH$_4$, SO$_2$, and CS$_2$, finding no robust evidence for any sulphur-bearing species despite the intense irradiation received by the planet. The inferred water abundance implies a strongly metal-enriched atmosphere. Elevated intrinsic temperatures can reconcile the non-detections of CH$_4$ and CS$_2$ through carbon--sulphur coupling, while the absence of SO$_2$ points toward a reduced atmospheric sulphur inventory or a near-solar to mildly enhanced C/O ratio, and places HD~106315~c near the transition between sulphur-rich and sulphur-poor chemical regimes.}
   {Our observations demonstrate the diversity of irradiated Neptunes beyond a simple description of the radiation field. HD~106315~c exhibits an atmospheric composition distinct from previously characterised warm Neptunes that show prominent sulphur photochemistry. The combination of a highly enriched atmosphere, tentative NH$_3$, and the absence of detectable sulphur-bearing species demonstrates that strong UV irradiation alone is not sufficient to guarantee observable sulphur photochemical products. These observations expand the diversity of atmospheric outcomes known for Neptune-mass exoplanets and provide new constraints on the coupled carbon, nitrogen, and sulphur chemistry operating in irradiated planetary atmospheres.}


   \maketitle

\section{Introduction}

Over the past three decades, exoplanet science has evolved from discovery to detailed atmospheric characterisation, from gas giants to smaller and cooler planets. Sub-Neptune and Neptune-sized planets, with radii between those of Earth and Neptune, are among the most common planetary populations in the Galaxy \citep{fulton_california-kepler_2017, bean_nature_2021}, yet their origins, internal structures, and atmospheric properties remain poorly understood. Atmospheric characterisation of transiting exoplanets provides one of the most powerful avenues for probing these questions, as atmospheric elemental abundances and molecular compositions encode information about planet formation, migration history, and subsequent atmospheric evolution \citep{mordasini_imprint_2016, molliere_interpreting_2022}. In particular, measurements of atmospheric metallicity and elemental ratios offer direct insight into the accretion history of planets within protoplanetary discs and the relative contributions of solids and gas during formation.

The advent of the James Webb Space Telescope (JWST) has transformed the study of exoplanet atmospheres. Its unprecedented spectral coverage and sensitivity now permit detailed spectroscopic investigations of Neptune-mass and sub-Neptune planets, enabling constraints on atmospheric composition, thermal structure, cloud properties, and chemical disequilibrium processes at a level previously inaccessible \citep[e.g.][]{madhusudhan_carbon-bearing_2023, benneke_jwst_2024, beatty_sulfur_2024, holmberg_possible_2024, hu_water-rich_2025, kreidberg_first_2025, davenport_toi-421_2025}. While early JWST atmospheric studies have predominantly focused on hot Jupiters and sub-Neptunes, Neptune-mass planets remain comparatively underexplored despite occupying a key transition regime between gas giants and smaller volatile-rich planets \citep{gressier_jwst-tst_2025}. Expanding the sample of well-characterised warm Neptunes is therefore essential for understanding the diversity of atmospheric compositions and chemical pathways across planetary mass regimes.

Prior to JWST, the Hubble Space Telescope (HST) and \textit{Spitzer Space Telescope} established the foundations of comparative exoplanet atmospheric studies. HST/WFC3 observations enabled robust detections of water vapour in numerous exoplanet atmospheres, while optical measurements from STIS and UVIS instruments constrained scattering processes and alkali absorption \citep[e.g.][]{deming_infrared_2013, sing_continuum_2016, tsiaras_population_2018,edwards_hst_pop,gascon_hustle_2025,Saba2025}. However, these facilities were largely restricted to the optical and near-infrared domains, limiting sensitivity to a relatively narrow set of molecular species and atmospheric processes. In particular, the absence of broad coverage at wavelengths beyond $\sim5~\mu$m prevented from obtaining robust constraints on atmospheric chemistry. As a consequence, atmospheric retrievals often suffered from degeneracies between molecular abundances, metallicity, and aerosol opacity \citep{constantinou_early_2023, espinoza_highlights_2025, k_barstow_computational_2026}.

One of the major advances enabled by JWST has been the emergence of sulphur chemistry as a new tracer of atmospheric photochemistry and metallicity. Sulphur-bearing molecules, particularly SO$_2$, CS$_2$ and H$_2$S, have now been detected in several highly irradiated exoplanet atmospheres using transmission spectroscopy \citep[e.g.][]{tsai_photochemically_2023, triantafillides_identification_2026, fu_hydrogen_2024}.
The detection of SO$_2$ in planets such as HAT-P-26b \citep{gressier_jwst-tst_2025}, WASP-39b \citep{alderson_early_2023,rustamkulov_early_2023, powell_sulfur_2024}, GJ~3470b \citep{beatty_sulfur_2024} and WASP-107~b \citep{dyrek_so2_2024, sing_warm_2024, welbanks_high_2024} has provided compelling evidence for photochemical processing driven by stellar ultraviolet (UV) irradiation. In many atmospheric models, SO$_2$ is produced through UV-driven oxidation of H$_2$S and becomes increasingly abundant at high metallicity and low C/O ratio \citep{polman_h2_2023}. However, recent theoretical work suggests that the abundance of sulphur-bearing species depends sensitively on temperature, irradiation environment, and elemental ratios, potentially leading to transitions between oxidised sulphur species such as SO$_2$ and reduced species such as CS$_2$ or H$_2$S \citep{Moses2024, de_gruijter_new_2025, veillet_inclusion_2026}. Understanding the conditions under which SO$_2$ and other sulphur bearing species are present or absent is therefore essential for interpreting atmospheric metallicities and chemical pathways in highly irradiated atmospheres.

Warm Neptune atmospheres are especially promising laboratories for studying these processes. Their extended atmospheres facilitate transmission spectroscopy, while their intermediate temperatures place them in a regime where both thermochemical equilibrium and photochemical disequilibrium can strongly influence observable compositions. In particular, the balance between nitrogen-, carbon-, and sulphur-bearing species is expected to be highly sensitive to vertical mixing, UV irradiation, and atmospheric metallicity. 

In this context, we introduce the Sulfur Understanding through Photochemistry and Exoplanetary Research (SUPER) program, a JWST Cycle II program (PID 2950, PI: R. Waters) which was designed to assess the presence of sulphur-bearing molecules in the atmospheres of three carefully selected exoplanets. These exoplanets, HAT-P-1~b, HAT-P-11~b and HD~106315~c, have a H/He-dominated atmosphere in which water was previously detected \citep{wakeford_hst_2013, fraine_water_2014, chachan_hubble_2019, guilluy_ares_2020, kreidberg_tentative_2022,edwards_hst_pop}, and span a wide range of temperatures. The main goal of the program is to establish the impact of different stellar radiation properties on the upper atmosphere photochemistry.

Our paper focuses on the third exoplanet in that list: HD~106315~c.
The HD~106315 multiplanet system was independently discovered by \cite{crossfield_two_2017} and \cite{rodriguez2017} based on observations from the K2 mission, and hosts two planets HD~106315~b and HD~106315~c, orbiting an F5V-type host star. HD~106315~c has a mass of $15.2 \pm 3.7 M_{\oplus}$ \citep{barros_precise_2017}, a radius of $4.379  \pm 0.086 R_{\oplus}$ \citep{howard_planet_2025} and a relatively low density of $0.65 \pm 0.23$ g cm$^{-3}$. The orbital period is $21.05731 \pm 0.00046$ days and was refined by \cite{lendl_ground-based_2017}. HD~106315~c has also a reported measurement of a sky-projected obliquity of $ \lambda =-2.68^{\circ +2.70^{\circ}}_{ -2.60^{\circ}}$ \citep{bourrier_dream_2023}. Its equilibrium temperature is estimated to be $886 \pm 20$ K \citep{barros_precise_2017}. The system also includes an inner, smaller companion planet, HD~106315b, with a radius of $2.18 \pm 0.33 R_{\oplus}$, which is most likely a ripped-core \citep{rodriguez2017}. Ground-based observations with the Keck helped assess that HD~106315~c is consistent with a $\sim$10\% H/He envelope and a possible Earth-like rock/iron core \citep{kosiarek_physical_2021}.

Orbiting a hot F-type star, HD~106315~c experiences a relatively strong UV irradiation environment that is expected to drive active photochemistry in its atmosphere. Near and mid-infrared observations are therefore uniquely suited to probing the resulting chemical composition, particularly for molecules such as H$_2$O, NH$_3$, CH$_4$, and sulphur-bearing species whose abundances are sensitive to both disequilibrium chemistry and atmospheric metallicity. In previous works, \cite{guilluy_ares_2020} and \cite{kreidberg_tentative_2022} analysed near-infrared observations of HD~106315~c taken with the WFC3 instrument on-board HST (PID 15333, PIs: I. Crossfield \& L. Kreidberg), covering a wavelength range from 1.088 to 1.68 $\mu$m. These analyses revealed the presence of an atmosphere with high statistical significance. Specifically, \citet{guilluy_ares_2020} retrieved a strong detection of H$_2$O with log$_{\rm 10}$[H$_2$O]$=-2.1^{+0.7}_{-1.3}$, and tentative evidence for NH$_3$ with log$_{\rm 10}$[NH$_3$]$=-4.3^{+0.7}_{-2.0}$, though the latter is not statistically significant. When ammonia is excluded from their model, the atmospheric model compensates by increasing the temperature to 1004 K and locating clouds at higher altitudes ($10^{2.5}$ Pa), which helps to suppress H$_2$O features and shrink the spectrum. Both works suggest two viable scenarios: a clear and highly metallic primary atmosphere, or a cloudy and hazy atmosphere with H$_2$O and traces of NH$_3$. In \cite{guilluy_ares_2020}, methane abundance is constrained only by an upper limit ($\sim10^{-5}$), while CO$_2$ and CO remain unconstrained.

Although HD~106315~c’s high equilibrium temperature would typically favour molecular nitrogen over ammonia, cooler temperatures at the terminator region (closer to 500 K) may still allow NH$_3$ to persist \citep{guilluy_ares_2020}. Since N$_2$ is spectrally inactive in the WFC3 range, it cannot be detected directly. \citet{guilluy_ares_2020} conducted several tests to assess the presence of N$_2$. Including it in their retrieval setup made no difference in the retrieved mean molecular weight, which led to inconclusive results. The combination of JWST spectroscopy with near-infrared observations offers a powerful opportunity to investigate the interplay between irradiation, photochemistry, and atmospheric structure in the atmosphere of this warm Neptune.

In this work, we present transmission spectroscopy of HD~106315~c obtained with the Mid-Infrared Instrument (MIRI) Low Resolution Spectrometer aboard JWST, combined with the archival HST/WFC3 observations described above, and contemporaneous UV and X-ray observations from \textit{Swift} and XMM-Newton. We independently reduce the JWST/MIRI, HST/WFC3, XMM-Newton and \textit{Swift} observations, extract the stellar energy distribution from the X-UV and associated data, perform atmospheric retrievals on the near and mid-infrared data using both the \texttt{TauREx} \citep{al-refaie_taurex_2021} and \texttt{POSEIDON} \citep{macdonald_hd_2017, macdonald_poseidon_2023} retrieval frameworks. We further interpret the observations using a grid of disequilibrium chemistry models computed with \texttt{VULCAN} \citep{tsai_vulcan_2017, tsai_comparative_2021} in order to investigate the role of metallicity, irradiation, and elemental abundance ratios in shaping the observed atmospheric composition. 

The paper is structured as follows. In Section~2, we describe the observations and data reduction. Section~3 presents the atmospheric retrieval analysis and inferred chemical abundances. In Section~4, we compare the observations with disequilibrium chemistry models. In Section 5, we discuss the implications for sulphur chemistry, methane depletion, atmospheric metallicity, and the formation history of HD~106315~c. Finally, Section ~6 summarises our conclusions and outlines prospects for future observations.
Appendices provide further detail on selected topics: A. Data analysis, B. Atmospheric free-chemistry retrieval, C. Atmospheric modelling and D. The stellar Spectral Energy Distribution (SED).

\section{Observations and Data reduction}

\subsection{JWST -- MIRI observations}
The observation consists of a mid-infrared transmission spectrum using the JWST MIRI Low Resolution Spectrometer \citep[LRS, eg.][]{kendrew_mid-infrared_2015} taken as part of the GO program 2950 (PI: R. Waters). The observational mode is slitless spectroscopy using the FASTR1 readout pattern. HD~106315~c was observed from June 13 2024 01:27:47 UTC to June 13 2024 12:51:48 UTC with a total duration of 11.28 hours and a transit duration of 4.64 hours. The data are split into two exposures, 16 uncalibrated files with a total of 6382 integrations of 19 groups each. No high-gain antenna movement was reported during the observation. The spectral coverage spans a range of 5 to 12 $\mu$m with a resolution of R$\sim$100. Data reduction is carried out with two independent pipelines and is described below. 

\begin{figure}[h!]
   \centering
   \includegraphics[width=\hsize]{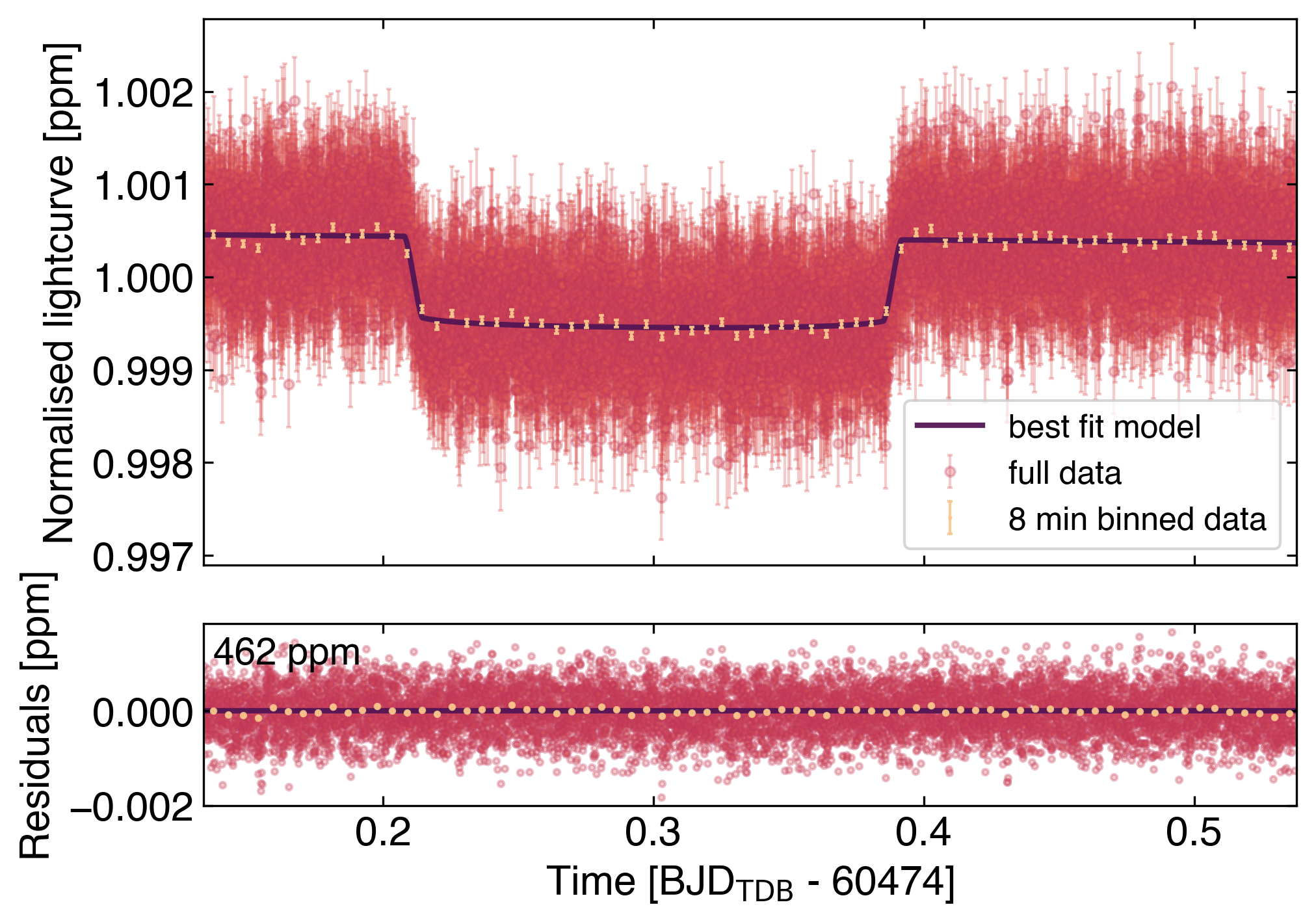}
      \caption{MIRI LRS white lightcurve of HD~106315~c based on the time series spectroscopic data binned between 4.9-12 $\mathrm{\mu}$m and discarding the first 1790 integrations. The top panel shows the white light curve, together with the median best-fit model. The lower panel shows the residuals after subtracting the model. The standard deviation of the unbinned residuals is 462 ppm.}
    \label{fig:white_fit}
\end{figure}

\begin{table*}[h!]
\caption{Parameters estimated from the white light curve analysis of the JWST MIRI LRS observation of HD~106315~c. The results
are shown from the Eureka! data analysis on a wavelength range between 4.9 and 12 $\mathrm{\mu}$m.}  
\label{table:white_parameters} 
\begingroup
\renewcommand{\arraystretch}{1.2}
\centering                   
    \begin{tabular*}{\textwidth}{@{\extracolsep{\fill}}lcc}
    \hline\hline              
    Parameter & Prior & Value \\       
    \multicolumn{3}{l}{Stellar parameters\tablefootmark{a}} \\
     Star radius [$R_{\odot}$] &  $\cdot \cdot \cdot $ & $1.269$ \\
     Star mass [$M_{\odot}$] &  $\cdot \cdot \cdot $ & $1.091$ \\
     Effective temperature [K] &  $\cdot \cdot \cdot $ & $6364$ \\
    \hline
    \multicolumn{3}{l}{Planetary fixed parameters} \\
    Period [days] & $\cdot \cdot \cdot$ & 21.056614 \\
    Limb-darkening coefficient $q_1$ & $\cdot \cdot \cdot$ & 0.209 \\
    Limb-darkening coefficient $q_2$ & $\cdot \cdot \cdot$ & 0.2115 \\
    Eccentricity & $\cdot \cdot \cdot$ & 0 \\
    \hline
    \multicolumn{3}{l}{Planetary retrieved parameters from white-light-curve fit} \\
    Mid-Transit time [BJD$_{\rm TDB}$] & $\mathcal{N}(60474.3, 0.05)$ & $60474.29989^{+0.00031}_{-0.00030}$ \\
    Inclination [$^{\circ}$] & $\mathcal{N}(89.4, 0.7)$ & $89.30^{+0.34}_{-0.29}$ \\
    Semi-major Axis [$a/R_{\mathrm{\star}}$] & $\mathcal{N}(33.4, 5)$ & $34.12^{+2.54}_{-2.73}$ \\
    Planet-to-star radius ratio [$R_{p}/R_{\mathrm{\star}}$] &  $\mathcal{U}(0.01, 0.05)$ & $0.03028^{+0.00025}_{-0.00026}$ \\
    GP log-amplitude  & $\mathcal{LU}(-25, -5)$ & $-18.027^{+0.119}_{-0.126}$ \\
    GP log-length scale & $\mathcal{LU}(-8, -3)$ & $-7.8914^{+0.1838}_{-0.1897}$ \\
    Out-of-transit flux [normalised] & $\mathcal{U}(0.99, 1.02)$ & $1.000376^{+9.45 \; 10^{-6}}_{-9.22 \; 10^{-6}}$ \\
    White noise [multiplicative factor] & $\mathcal{U}(1, 10)$ & $2.472^{+0.0179}_{-0.0169}$ \\
    \hline
    \multicolumn{3}{l}{Planetary derived parameters} \\
    Planet radius [$R_{J}$] &  $\cdot \cdot \cdot $ & $0.3738^{+0.00316}_{-0.00313}$ \\
    Semi-major axis [AU] &  $\cdot \cdot \cdot $ & $0.2013^{+0.1846}_{-0.2163}$ \\
    \hline
    \end{tabular*}
    \tablefoottext{a}{The stellar radius and effective temperature are taken from \cite{howard_planet_2025}, and the stellar mass comes from \cite{barros_precise_2017}.}
\endgroup
\end{table*}


\subsubsection{Eureka! pipeline}\label{section:eureka}

First, we perform Stages 1 and 2 using the \texttt{jwst} calibration pipeline \citep{bushouse_jwst_2025} to perform the detector ramp-fitting and instrument-level calibration (version 1.17.1, pmap 1322). In Stage 1, we run the data quality initialization (DQ), correction from contaminating electromagnetic frequencies (EMIcorr), saturation flagging, last frame flagging, non-linearity correction using a custom non-linearity correction file (see \citet{dyrek_so2_2024} and Appendix~\ref{section:custom linearity}), reset switch charge decay (RSCD) flagging, dark current subtraction and reset correction, cosmic ray jump detection with a rejection threshold of 5$\sigma$ and ramp fitting. No RSCD-flagging produces very similar results, as shown in Figure~\ref{fig:spectrum_rscd}. 
In Stage 2, we perform flat-field correction. Next, for each integration, we subtract the background by taking the median flux of columns outside the trace, from 12 to 20 and from 52 to 60.
We run Stages 3--5 using the \texttt{Eureka!} pipeline \citep{bell_eureka_2022}. We extract the spectra using an optimised Gaussian PSF extraction with an aperture of 8 pixels. We choose a bin size of 0.2 $\mu$m to extract the spectral lightcurves, and outliers are masked using a running median sigma-clipping method with a rejection threshold of 5$\sigma$ and a box size of 20 data points. The choice of the bin size is optimised to mitigate additional noise in the final transmission spectrum \citep[e.g.][]{bell_nightside_2024}. We perform lightcurve fitting using the \texttt{batman} \citep{kreidberg_batman_2015} transit model and the nested sampling Bayesian inference framework with the \texttt{dynesty} sampler \citep{speagle_dynesty_2020}. To infer the orbital parameters, we first perform a white-lightcurve fit summing all wavelengths from 4.9 to 12 $\mu$m. The argument of periastron is set to 90 degrees with a circular orbit. The period is fixed to $P = 21.056614$ days \citep{kokori_exoclock_2023}. We fit for the normalised semi-major axis $a/\mathrm{R}_{\star}$, the inclination $i$, the radius $R_p/\mathrm{R_{\star}}$ and the transit mid-time $t_0$ using wide Normal priors based on values reported in the literature \citep{kokori_exoclock_2023}. The systematics are fitted using a linear trend $c_0 F_{\rm obs}(t) + c1$ where $c_0$ and $c_1$ are the systematics coefficients and the normalized $F_{\rm obs}(t)$ is the flux. We also fit for a white noise inflation factor. The fit is performed after trimming 1790 integrations (1.5 hours) to remove the detector settling exponential decay. The limb darkening coefficients are fixed to values retrieved from the \texttt{ExoCTK} tool \citep{bourque_exoplanet_2021} and converted following the \cite{kipping_efficient_2013} parametrisation ($q_1 = 0.209$ and $q_2 = 0.2115$). Fitting for $q_1$ and $q_2$, or fixing them to wavelength-dependent values, shows very similar results in the obtained final spectrum. The white lightcurve fit is shown in Figure~\ref{fig:white_fit} and the parameters retrieved from the fit are presented in Table~\ref{table:white_parameters}. 
The spectroscopic fit is performed fixing the normalised semi-major axis $a/\mathrm{R}_{\star}$, the inclination $i$, the transit mid-time $t_0$, the orbital period $P$ and the limb darkening coefficients $q_1$ and $q_2$ retrieved from the white lightcurve fit. The fitted parameters are the radius $R_p/\mathrm{R_{\star}}$, the systematics model similar to the white lightcurve fit and a white noise inflation factor. We also explore the sensitivity of the retrieved spectrum to a different bin size of 0.4 $\mu$m in Appendix~\ref{section:spectral_binning}. The retrieved spectrum is shown in Figure~\ref{fig:spectrum_binsize}. The residuals of the fit show evidence for time-correlated noise. The timescale of the time-correlated noise is $\sim$1 minute, and the amplitude of the correlation increases at shorter wavelengths, as discussed in Appendix~\ref{section:time-correlated noise}, consistent with previous work \citep{Madhusudhan2025}. Figure~\ref{fig:correlated_noise} shows the correlation timescales and amplitudes. To mitigate the issue, we inject Gaussian Processes (GPs) into the fit. As a consequence, the uncertainties of the spectral points increase up to 50$\%$ at 4.9 $\mu$m.

\subsubsection{JExoRES pipeline}\label{section:jexores}

We use a second pipeline called \texttt{JExoRES} \citep{holmberg_exoplanet_2023}, which was recently adapted for MIRI LRS \citep{Madhusudhan2025}, for additional robustness. Stages 1 and 2 are similar to Sect.~\ref{section:eureka} and use the \texttt{jwst} calibration pipeline \citep{bushouse_jwst_2025}. The following steps are applied to the data: DQ initialisation, EMI correction, saturation flagging, standard linearity correction, RSCD flagging, dark current subtraction, and ramp fitting. We then apply the flat-field correction. In Stage 3 of the \texttt{JExoRES} pipeline, we first use the gain reference file to convert the flux from DN s$^{-1}$ to e$^{-}$ s$^{-1}$. We then search for cosmic-ray hits by performing sigma clipping on the time series of each pixel by first removing a running median of 7 integrations and using a threshold of $7\sigma$. In addition, we also mask neighbouring pixels surrounding detected outliers as well as pixels that have been flagged with any issues during Stages 1–2. We then correct for the background by subtracting the mean of the flux outside the trace, for each detector column and integration. To estimate the background, we use columns with pixel numbers 13-29 and 43-59, while not including bad pixels or outliers. Finally, we extract the spectra using optimal extraction \citep{Horne1986}, using the median of all integrations as a model for the point-spread function for each spectral channel. For this, we use an aperture of 9 pixels. 

Next, we fit the light curves using the \texttt{batman} \citep{kreidberg_batman_2015} transit model and perform nested sampling with \texttt{MultiNest} \citep{buchner_johannesbuchnermultinest_2021}, setting the eccentricity to 0 and the period to the value from \citet{kokori_exoclock_2023}, similar to Sect.~\ref{section:eureka}. We perform the spectroscopic light-curve fitting by fixing $a/\mathrm{R}_{\star}$, $i$, $t_0$, and the two limb-darkening coefficients to the values from Table~\ref{table:white_parameters}. We fit for $R_p/\mathrm{R_{\star}}$, the systematics model parameters and a white noise inflation factor. Limb-darkening coefficients are fixed to the same values as described in Sect.~ \ref{section:eureka}. We bin the light curves in wavelength prior to fitting, with a width of 0.2~$\mu$m or a minimum of 4 pixels, same as in Sect.~\ref{section:eureka}. For the baseline trend, we adopt a linear trend and discard 1790 integrations to mitigate the effects of the detector settling ramp, as in Sect.~\ref{section:eureka}. The error bars are also inflated to account for time-correlated noise using GPs \citep{Madhusudhan2025}.

\subsection{Robustness of the transmission spectrum}\label{section:robustness}

We explore the sensitivity of the spectrum to the following inputs: the \texttt{jwst} pipeline RSCD step that masks the first four frames of the ramp at the Stage 1 level, and the spectral binning. All results are presented in Appendix~\ref{section:reduction comparison}. The two spectra from different data reductions are consistent within 1$\sigma$ error apart from a data point at 10.8 $\mu$m. 
All tests run on the dataset to perform sanity checks provide spectra compliant within 1$\sigma$ as well. This way, we ensure the robustness of the retrieved spectrum and no dependency on the different treatments of the data: non-linearity correction, different numbers of frames used for ramp-fitting or different systematics trends. The 0.4 $\mu$m binned spectrum in Figure~\ref{fig:spectrum_binsize} however, indicates a possible bias in the retrieved spectral point at 8.5 $\mu$m. Investigations of the corresponding pixels' lightcurves show a jump in flux right after transit, possibly due to a cosmic ray hit. The corresponding pixels do not recover over time, and the flux stays higher (see Appendix~\ref{section:time-correlated noise}). As a consequence of the baseline change, the fit leads to a low spectral feature (around $\sim$600 ppm). 
When conducting atmospheric retrievals with all three spectra, we run sanity tests by removing either the first spectral point, or the one at 8.5 $\mu$m, or both from the spectrum, to avoid any bias from instrumental systematics.

\subsection{HST--WFC3 observations}

Four transits of HD~106315~c were observed on November 3 and December 21-22, 2018 and then on February 2 and November 21, 2019. Observations were obtained with the WFC3 on board HST using the 1.41 µm Grism (G141), spanning a range from 1.088 to 1.68 $\mu$m. The data were obtained as part of the GO program 15333 (PIs: I. Crossfield \& L. Kreidberg). Two independent data analyses were conducted by \cite{guilluy_ares_2020} and \cite{kreidberg_tentative_2022}. For consistency, we perform an independent calibration and light curve fitting of the HST data using the \texttt{CASCADe} package \citep[e.g. ][]{lahuis_horizon_2020,carone_indications_2021}, using the same orbital and stellar parameters as used for the analysis of the JWST MIRI lightcurve data. Before fitting, the spectral lightcurves are binned from the original spectral resolution to a uniform wavelength grid with a spectral bin width of 0.00757 $\mu$m. Additional regression parameters to the systematics model \citep[for more details, see ][]{carone_indications_2021} are the time variable and the trace position. The derived band-averaged transit depth is 0.0303315 $\pm$ 0.003 $R_{p}/R_{\mathrm{\star}}$, consistent within 1$\sigma$ with the transit depth derived from the JWST MIRI observations. A bootstrap analysis is then used to estimate the error bars on the spectral points, similarly to the method used to derive the error on the white transit depth. All three spectra are in agreement within 1$\sigma$.

\section{The stellar spectrum}

Most well-characterised Sub-Neptunes and Neptune-sized exoplanets orbit cool M- or K-type stars \citep{mulders_stellar-mass-dependent_2015}. HD~106315 is an F-type star and the hottest one in our sample. As such, planets around F-type stars may experience irradiation conditions that differ fundamentally from those of cooler hosts. Their continuous
UV fields drive persistent photochemistry and atmospheric escape, unlike the flare-dominated and time-variable UV environments of M dwarfs \citep[see e.g.][]{france_muscles_2016, Linsky2019, Behr2023}.
Measurements of short-wavelength emissions in the X through the near-UV (NUV) of the host star ($ \sim 1$--3000 \AA) allow us to assess the strength of the irradiation field received by HD~106315~c and its impact on the atmospheric chemistry. 

We report here the results of NUV and X-ray observations with the Neil Gehrels Swift Observatory \citep[\textit{Swift},][]{Gehrels2004}, and the XMM-Newton Observatory \citep{Jansen2001}, performed specifically in association with the JWST/MIRI observations of HD~106315~c.
Prior to these observations, the only other measurements at wavelengths shorter than the visible appear to be from the GALEX all-sky UV survey conducted from 2003 to 2012 \citep{loyd2020}.
XMM-Newton monitored HD~106315 from June 19 2024 $\sim$09:45~UTC to June 20 2024 $\sim$ 01:06~UTC (PID 0940590201), only a few days after the JWST/MIRI observations, for a total duration of $\sim 38$ hours in a single, continuous exposure.
\textit{Swift} monitored HD~106315 during part of the JWST/MIRI observations of HD~106315~c from June 13 2024 00:28 UTC to 02:30 UTC, and during part of the XMM-Newton observation from
June 19 2024 23:52 UTC to June 20 2024 00:22 UTC, for a 33-minute exposure time on each occasion (PID 16660), which provides a link between the JWST and XMM-Newton datasets.



\subsection{XMM-Newton X-ray observations} \label{sec:stellar_xray}

We reduce the observing data set following standard procedures as described in the XMM-Newton data analysis threads\footnote{\url{https://www.cosmos.esa.int/web/xmm-newton/sas-threads}}. Important steps include running the pipelines for the three European Photon Imaging Cameras (EPIC cameras) applying calibration steps, checking for intervals with high proton background levels (no bad intervals were identified as the background level was very low throughout the observation), extracting the source counts and counts from a low-level background using circular regions on each camera to produce binned low-resolution X-ray spectra, and finally creating the response matrix files as well as ancillary response files. Given the faint source, our spectral bins contained at least 15 or 25 counts each. Among the three cameras, we found that only the pn-CCD camera produced a spectrum of sufficient quality to allow for a quantitatively meaningful spectral interpretation.

We fit the pn-CCD spectrum using the \texttt{XSPEC} software \citep{arnaud1996} and bin the data to at least 25 counts per bin to get a more reliable fit. The spectrum is modelled using a one-component, isothermal, optically thin coronal collisional ionisation equilibrium \textit{vapec}-type plasma, which allows us to set element abundances relative to the abundances of the solar photosphere. We use a set of abundances reported in \citet{guedel2007} representing an average over a wide, well-observed sample of magnetically active stars.  Our results show a plasma electron temperature of 2.2~MK and an emission measurement that corresponds to a total X-ray luminosity of $2.46\times 10^{27}$~erg~s$^{-1}$ over the 0.1--10~keV energy band, for a distance of 109~pc from the source. These values closely correspond to solar values during the Sun's activity maximum in its 11-year activity cycle \citep{peres2000}.
More details on the XMM-Newton dataset analysis are provided in Appendix~\ref{app:stellar_sed}.



\subsection{XMM-Newton and \textit{Swift} NUV observations} \label{sec:stellar_nuv}

\paragraph{XMM-Newton/OM photometry and spectroscopy.} \label{sec:stellar_nuv_xmm}


From the XMM Science Archive (XSA)\footnote {\url{https://nxsa.esac.esa.int/nxsa-web/\#home}}, we download the standard data products provided by the pipeline processing system (PPS)\footnote{\url{https://www.cosmos.esa.int/web/xmm-newton/pipeline}}, and use primarily the source lists and calibrated the extracted grism spectra.
We detect the target star at high significance in the two photometric NUV filter bands, and spectra are also obtained with the UV-grism. Table~\ref{tab:nuv_results} summarises the mean flux density in each photometric bandpass, the corresponding luminosity, and the flux density incident on HD~106315~c.

We combine the two individual grism exposures over the full range of 1800--3600 $\mathring{\rm A}$ by simple averaging, as they had similar exposure times. The grism data shows artefacts due to instrumental calibration issues \citep[e.g. the "Jupiter patch",][]{sullivan2023} and possible contamination from nearby, weaker but partially overlapping spectra. In our case, the wavelength range from $\sim 2100$\ \AA\ to $\sim 3000$\ \AA\ shows no obvious problems. 

\begin{figure}
   \centering
   \includegraphics[width=\columnwidth]{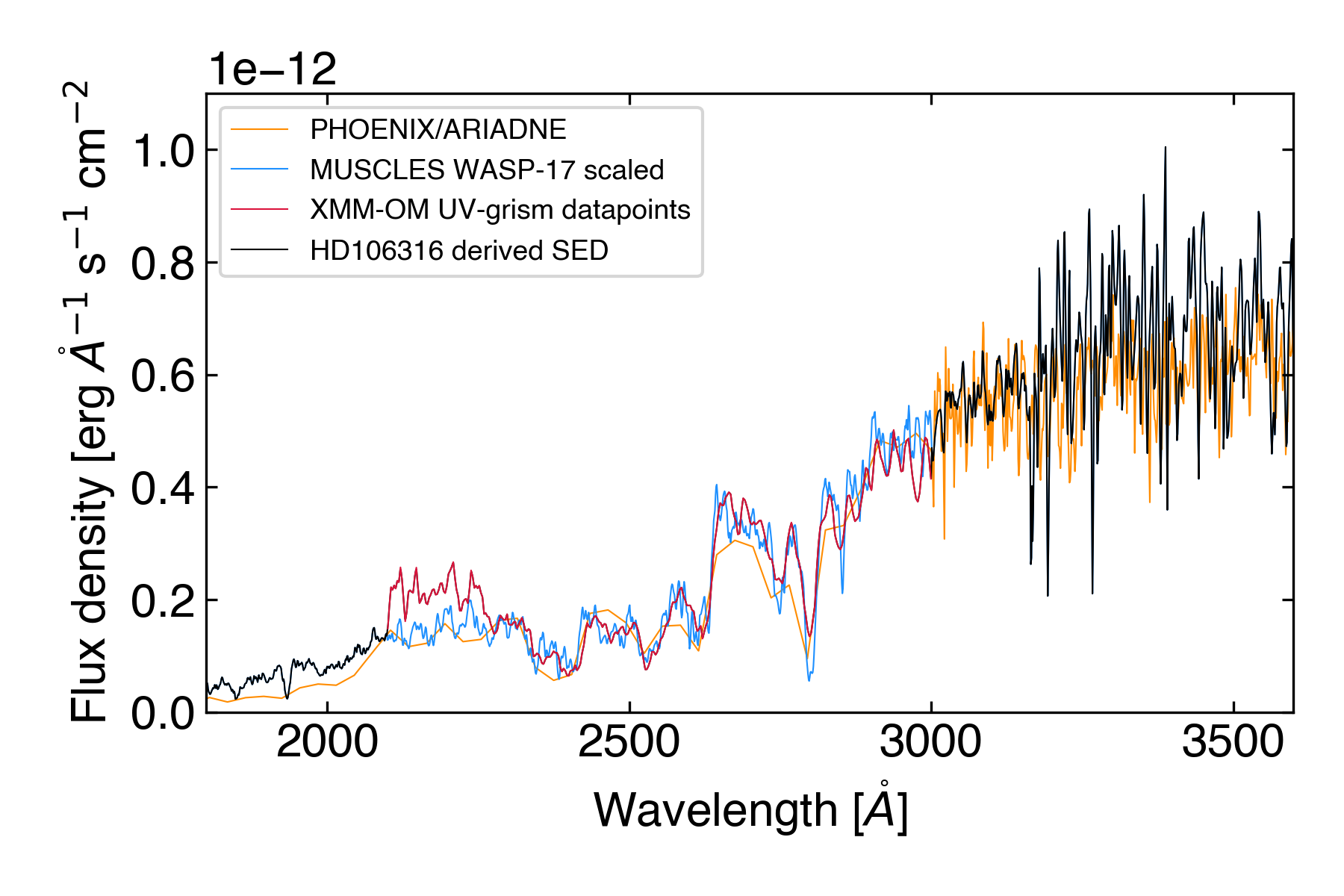}
      \caption{XMM-Newton/OM UV-grism spectrum of HD~106315; this is the mean of the two exposures.  
      The values are flux- and wavelength-calibrated, and background-subtracted -- see Sect.~\ref{sec:stellar_nuv_xmm} for details.
      The scaled MUSCLES spectrum of WASP-17 \citep{Behr2023}, the
      PHOENIX model spectrum, and our derived SED, are shown for comparison.
      (The derived SED uses the XMM-OM UV-grism spectrum over the wavelength range 
      2100--3000\ \AA.)
      }
    \label{fig:xmm_nuv_grism}
\end{figure}

We show, in Fig.~\ref{fig:xmm_nuv_grism}, the comparison between the grism spectra, the MUSCLES WASP-17 SED, which has a similar spectral type and age/activity-level as HD~106315 \citep{france_muscles_2016, Behr2023} scaled in flux, and a PHOENIX model atmosphere fitted to optical and near-IR photometry (see Appendix~\ref{app:stellar_sed}).
The scaling of the MUSCLES spectrum is also described in Appendix~\ref{app:stellar_sed}.
The XMM spectra are shifted in wavelength by +27~\AA\ to align with the PHOENIX model and the WASP-17 spectra, based on obvious features in the range $\sim 2400$--2900~\AA. Within the range from $\sim 2100$\ \AA\ to $\sim 3000$\ \AA, all the spectra show generally good agreement in flux.

\paragraph{\textit{Swift}/UVOT photometry.} \label{sec:stellar_nuv_swift}

Similarly to the XMM-Newton/OM photometric data, we obtain two \textit{Swift}/UVOT photometric datasets with the uvm2 filter, centred on $2246 \pm 498$ \AA\ \citep{poole2008}. For these datasets, the observational configuration and data processing scheme are similar to those described in \citet{Dyrek2024} (Supplementary Information, Sect.~2.1), with the processing done using the \textit{Swift} software tools. Both of our observations yields high-significance detections of the star. All results are summarised in Table~\ref{tab:nuv_results}. 




\begin{table*}
\renewcommand{\arraystretch}{1.25} 
\centering
\caption{XMM-Newton and \textit{Swift} NUV photometry of HD~106315. }
\begin{tabular}{llrrrr}
\hline \hline
Instrument/ & Observation & Flux density (at Earth)                  & Luminosity                     & Flux density at HD~106315~c               & Relative  \\ 
filter      & date$^a$        & $ (\rm 10^{-13}\ erg\ cm^{-2}\ s^{-1}\ \AA^{-1} $) & ($\rm 10^{29}\ erg\ s^{-1}\ \AA^{-1}) $ & ($\rm 10^{3}\ erg\ cm^{-2}\ s^{-1}\ \AA^{-1} $) & error$^b$ (\%) \\
\hline 
XMM OM uvm2 & 2024-06-19      & 1.59                                       & 2.26                         & 3.57                                  & 0.1   \\
XMM OM uvw2 & 2024-06-19      & 1.52                                       & 2.16                         & 3.41                                  & 0.4   \\
Swift UVOT uvm2 & 2024-06-19  & 1.55                                       & 2.20                         & 3.48                                  & 2   \\
Swift UVOT uvm2 & 2024-06-13  & 1.59                                       & 2.26                         & 3.57                                  & 2   \\
\hline
\end{tabular}
\label{tab:nuv_results}
\tablefoot{(a) year-month-day; (b) 1 standard deviation statistical error }
\end{table*}

\subsection{The final stellar SED}\label{sec:stellar_sed}

Using all the results presented above and in Appendix~\ref{app:stellar_sed}, we then assessed the stellar Spectral Energy Distribution (SED) following the methodology of the MUSCLES project \citep{france_muscles_2016, Behr2023}, therefore covering the full wavelength range from $\sim 5$ to $\sim 50000$ \AA. The composite SED is shown in Fig.~\ref{fig:stellar_sed} and further details of the SED and its construction are given in Appendix~\ref{app:stellar_sed}.
This SED allows us to characterise the wavelength-dependent irradiation received by HD~106315~c. For photochemistry, the spectral shape of the stellar flux, and in particular the relative NUV and FUV contributions, is more important than the total bolometric irradiation received by the planets. We discuss such implications for the atmospheric chemistry of HD~106315~c in Sect.~\ref{sec:discussion_sulphur}.


\section{Atmospheric Retrievals}\label{sect:atm_retrievals}

\subsection{Retrieval setups}

We conduct free-chemistry retrievals with the inclusion of a grey cloud deck on both the HST and JWST MIRI LRS observations of the transmission spectrum of HD~106315~c using two open-source atmospheric retrieval codes \texttt{POSEIDON} \citep{macdonald_hd_2017, macdonald_poseidon_2023}, and \texttt{TauREx} \citep{al-refaie_taurex_2021}. We use two independent retrieval frameworks to assess the robustness of our results to the retrieval implementation, as \texttt{POSEIDON} and \texttt{TauREx} employ independent numerical and radiative-transfer implementations.

For both sets of retrievals, we assume that HD~106315~c possesses a primordial helium/hydrogen-dominated atmosphere with a solar helium-to-hydrogen number density (He/H$_2$ = 0.1765). We employ an isothermal pressure-temperature ($P-T$) profile, setting bounds of $T \in [200, 1500]$~K. The temperature in the retrieval is constrained from its influence on the scale height (and thus on the amplitude of the transit spectral features) and the spectral shape of the molecular bands. We also consider non-isothermal $P-T$ profiles, but they don't influence the inferences about the atmosphere. It has indeed been shown that isothermal $P-T$ profiles are sufficient in most transmission geometries \cite{macdonald_why_2020}. Free-retrievals also prevent the atmospheric composition from being biased by the non-isothermal structure of the deep atmosphere. We assume a simple grey-cloud model, where the only fitting parameter is the cloud pressure (CP) that has bounds of CP $\in [10^{-10}, 10^2]$~bar. Further exploration of different types of clouds and hazes is presented in Sect.~\ref{sect:clouds}. Stellar parameters (stellar radius, effective temperature, stellar metallicity and surface gravity in the log-space) are fixed to the following values \citep{howard_planet_2025}: R$_\star$ = 1.269 R$_\odot$, T$_{\rm eff}$ = 6364 K, Met$_{\star}$ = -0.22 dex, $\log$ g$_\star$ = 4.291 cm s$^{-2}$, while the planetary radius is fitted between 0.2 and 0.5 $\rm{R_{Jup}}$.

\subsection{Atmospheric retrievals with \texttt{TauREx}}

The atmosphere is modelled between $P \in [10^{-10}, 10^2]$~bar using 100 plane-parallel layers uniformly partitioned in log-space. We include collision-induced absorption from H$_2$-H$_2$ \citep{abel_h2-h2, fletcher_h2-h2} and H$_2$-He \citep{abel_h2-he} as well as Rayleigh scattering for all molecules. For each molecule, we set bounds on the volume mixing ratio (VMR) of $\rm log_{10}(VMR) \in [-15, -1]$, assuming a constant abundance with altitude. We use correlated k-tables to compute the opacities, and all opacities were taken from the ExoMol database\footnote{\url{https://www.exomol.com/data/molecules/}} \citep{chubb_database}. We explore the parameter space using the nested sampling algorithm \texttt{MultiNest} \citep{Feroz_multinest,buchner_johannesbuchnermultinest_2021} with 1000 live points and an evidence tolerance of 0.5, which is the value recommended in the documentation\footnote{\url{https://github.com/JohannesBuchner/MultiNest/blob/master/README}}. We fit for an offset between the two datasets, allowing the MIRI data points to vary with an overall offset between -100 and 100 ppm. For molecular absorptions, we include H$_2$O \citep{polyansky_h2o}, CH$_4$ \citep{yurchenko_ch4}, NH$_3$ \citep{coles_exomol_2019}, HCN \citep{barber_exomol_2014},
CO \citep{li_co}, CO$_2$ \citep{yurchenko_co2}, H$_2$S \citep{azzam_h2s}, SO$_2$ \citep{underwood_so2} and OCS \citep{owens_ocs}.

\subsection{Atmospheric retrievals with \texttt{POSEIDON}}

The \texttt{POSEIDON} retrievals \citep{macdonald_hd_2017, macdonald_poseidon_2023} adopt the same model and prior distribution configuration described above. For each molecule, priors on log10(VMR) are set to  $[-12, -1]$, assuming a constant abundance with altitude. We use a similar number of live points and evidence tolerance. The reference pressure is set to 10 bar, the retrieved reference radius being the radius at this pressure. We fit for an offset between the two datasets similarly to the \texttt{TauREx} retrieval, allowing the MIRI data points to vary relatively to the WFC3 with an offset between -1000 and 1000 ppm. For molecular absorptions, we include H$_2$O \citep{polyansky_h2o}, CH$_4$ \citep{yurchenko_ch4}, NH$_3$ \citep{coles_exomol_2019}, HCN \citep{barber_exomol_2014},
CO \citep{li_co}, CO$_2$ \citep{yurchenko_co2}, H$_2$S \citep{azzam_h2s}, SO$_2$ \citep{underwood_so2}, OCS \citep{owens_ocs} and CS$_2$ from the HITRAN database \citep{gordon_hitran2020_2022}.

\subsection{Atmospheric retrievals results}\label{section:retrievals}

\begin{figure*}
    \centering
    \includegraphics[width=\linewidth]{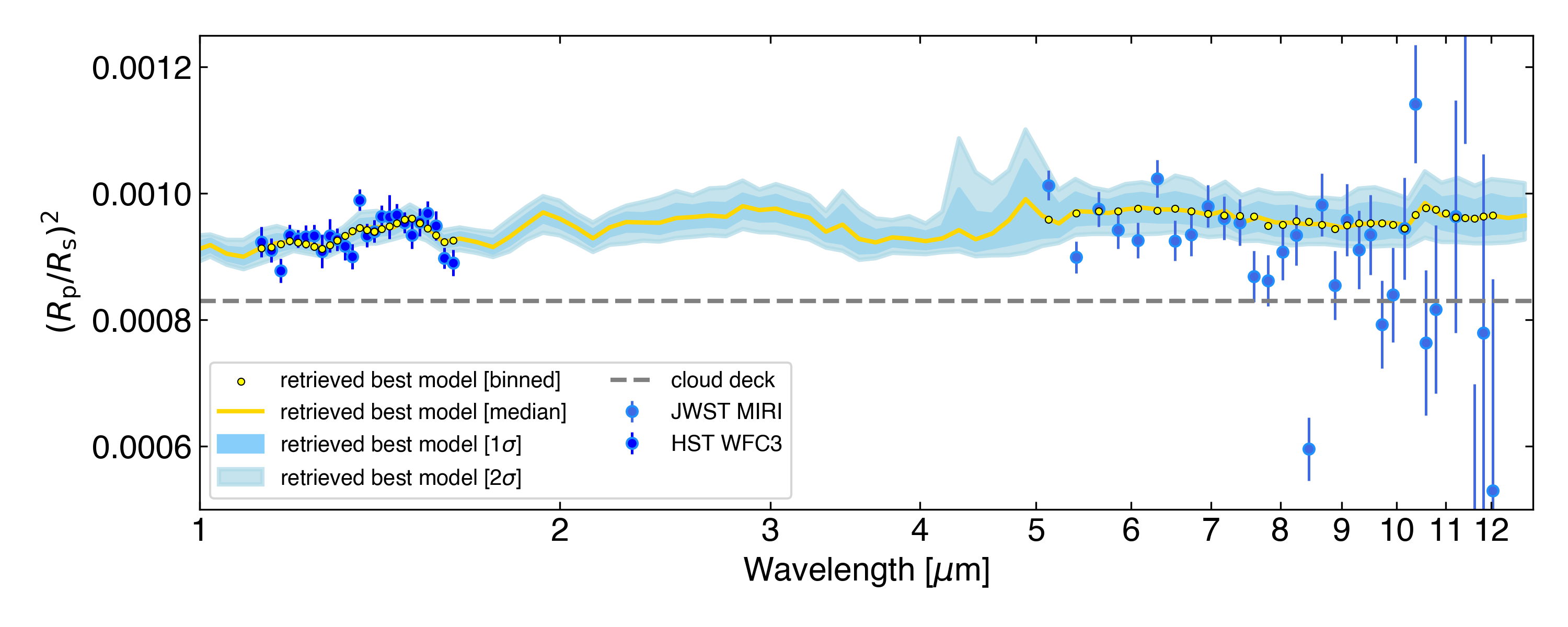}
    \caption{Best fit model from atmospheric retrievals and comparison with the HST/WFC3 and JWST/MIRI observations. A retrieved mean offset of -2.91 ppm is applied to the MIRI data. The $\sim 600$ ppm data point at 8.5 $\mu$m is not included in the fiducial retrieval but is displayed here for completeness.}
    \label{fig:best_fit_baseline_spec}
\end{figure*}

The fiducial spectra used to perform the main retrievals are the JWST MIRI \texttt{Eureka!} baseline reduction without the data point at 8.5 $\mu$m (see Sect.~\ref{section:robustness}), and the HST WFC3 \texttt{CASCADe} reduction obtained through this work. The best-fit atmospheric model retrieved with \texttt{POSEIDON} to the fiducial spectra is shown in Figure \ref{fig:best_fit_baseline_spec}. 
To quantify the significance of the detections, we perform the standard Bayesian model comparison used for molecular detections, as in \cite{benneke_how_2013}, by running retrievals removing each species one by one. We then compute the preference for their presence using the difference in the log Bayesian evidence $\Delta \ln Z$. We report the difference in log Bayesian evidence rather than converting it into a Gaussian-equivalent detection significance, as different prescriptions for such a conversion can lead to substantially different values \citep{sellke_calibration_2001, trotta_bayes_2008, benneke_how_2013,kipping_exoplaneteers_2025}. 
Overall, the atmospheric retrievals consistently indicate the presence of H$_2$O ($\Delta \ln Z$ = 6.25) and NH$_3$ ($\Delta \ln Z$ = 4.16) in the atmosphere of HD~106315~c. We retrieve a water abundance of $\log_{10}(\mathrm{VMR})=-1.40^{+0.40}_{-0.74}$ and an ammonia abundance of $\log_{10}(\mathrm{VMR})=-2.47^{+0.57}_{-0.89}$, in agreement with previous analyses of the HST/WFC3 transmission spectrum \citep{guilluy_ares_2020,edwards_hst_pop}. The retrieved isothermal temperature of $747^{+150}_{-155}$ K is consistent within the uncertainties with the planet's equilibrium temperature of $\sim$886 K. The cloud deck is constrained to pressures of $\log_{10}(P_{\rm cloud}/{\rm bar}) = -0.41^{+1.50}_{-1.64}$, indicating that any cloud opacity is located relatively deep in the atmosphere, with the marginal posterior distribution pushed against the upper limit of the prior.

For all other molecules considered in the retrievals, only upper limits are obtained. In particular, we find no evidence for CH$_4$, SO$_2$, CS$_2$, CO, CO$_2$, HCN, H$_2$S, or other sulphur-bearing species. For CH$_4$, SO$_2$, and CS$_2$, we derive 2$\sigma$ upper limits of $\log_{10}(\mathrm{VMR})<-5.86$, $\log_{10}(\mathrm{VMR})<-6.28$, and $\log_{10}(\mathrm{VMR})<-3.60$, respectively. The probability distributions for these are shown in Figure \ref{fig:mol_constraints}. The absence of detectable SO$_2$ is particularly noteworthy given the strong irradiation received by the planet from its F-type host star and the increasing number of SO$_2$ detections reported in JWST transmission spectra of warm exoplanets. Equally intriguing is the lack of detectable CH$_4$. At the retrieved atmospheric temperature of $\sim$700\,K, thermochemical and photochemical models generally predict methane to be one of the dominant carbon-bearing species \citep{moses_compositional_2013}. The stringent upper limit derived here therefore suggests that methane is significantly depleted relative to these expectations and that additional atmospheric processes are modifying the chemical composition of the observable atmosphere.

Although some retrieval configurations show a preference for OCS ($\Delta \ln Z = 1.17$ for our fiducial retrieval) with an upper limit of the probability distribution near $\log_{10}(\mathrm{VMR})\sim -2.20$, this result is not robust across reductions and appears to be driven primarily by the shortest-wavelength MIRI data point. We therefore do not consider OCS to be tentatively detected. Instead, the most robust conclusion from the retrieval analysis is the combination of significant H$_2$O and NH$_3$ abundances, a relatively cool atmospheric temperature, and stringent upper limits on CH$_4$, SO$_2$, and CS$_2$. These constraints provide a valuable benchmark for interpreting the disequilibrium carbon, nitrogen, and sulphur chemistry of HD~106315~c.

\begin{table}[h!]
\caption{Retrieved abundance estimates from JWST MIRI LRS and HST WFC3 observations of HD~106315~c.}  
\label{table:retrieval_abundance} 
\begingroup
\renewcommand{\arraystretch}{1.2}
\centering                   
\begin{tabular*}{\columnwidth}{@{\extracolsep{\fill}}lccc}
\hline\hline              
Parameter &  Prior & Value & $\Delta \ln Z$\tablefootmark{a} \\ 
\multicolumn{4}{l}{Retrieved} \\
$R_p$ [$R_{\rm Jup}$] & $\mathcal{U}(0.2, 0.6)$ & $0.35^{+0.01}_{-0.01}$ & $\cdot \cdot \cdot$\\
$T$ [K] & $\mathcal{U}(100,1500)$ & $747^{+150}_{-155}$ & $\cdot \cdot \cdot$\\
$\log$ H$_2$O & $\mathcal{U}(-1,-12)$ & $-1.40^{+0.40}_{-0.74}$ & 6.25\\
$\log$ NH$_3$ & $\mathcal{U}(-1,-12)$ & $-2.47^{+0.57}_{-0.89}$ & 4.16\\
$\log$ CH$_4$ & $\mathcal{U}(-1,-12)$ & $<-5.86$ & $\cdot \cdot \cdot$\\
$\log$ SO$_2$ & $\mathcal{U}(-1,-12)$ & $<-6.28$ & $\cdot \cdot \cdot$\\
$\log$ CS$_2$ & $\mathcal{U}(-1,-12)$ & $<-3.60$ & $\cdot \cdot \cdot$\\
$\log$ OCS & $\mathcal{U}(-1,-12)$ & $<-2.20$ & 1.17\\
$\log$ CO & $\mathcal{U}(-1,-12)$ & $<-2.44$ & $\cdot \cdot \cdot$\\
$\log$ CO$_2$ & $\mathcal{U}(-1,-12)$ & $<-3.51$ & $\cdot \cdot \cdot$\\
$\log$ HCN & $\mathcal{U}(-1,-12)$ & $<-3.91$ & $\cdot \cdot \cdot$\\
$\log$ H$_2$S & $\mathcal{U}(-1,-12)$ & $<-4.42$ & $\cdot \cdot \cdot$\\
Offset$_{\rm HST/WFC3}$ & $\mathcal{U}(-10^{3},10^{3})$ &
$-2.91^{+1.78}_{-2.04}$ & $\cdot \cdot \cdot$\\
$\log P_{\rm clouds}$ [bar] & $\mathcal{U}(-6,2)$ &
$-0.41^{+1.50}_{-1.64}$ & $\cdot \cdot \cdot$\\
\hline
\multicolumn{4}{l}{Derived} \\
MMW & $\cdot \cdot \cdot$ & $3.30^{+1.29}_{-0.72}$ & $\cdot \cdot \cdot$ \\
log$_{\rm 10}$(O/H) & $\cdot \cdot \cdot$ & $-1.33^{+0.38}_{-0.58}$ & $\cdot \cdot \cdot$ \\
log$_{\rm 10}$(C/H) & $\cdot \cdot \cdot$ & $-2.75^{+0.84}_{-1.32}$ & $\cdot \cdot \cdot$ \\
log$_{\rm 10}$(N/H) & $\cdot \cdot \cdot$ & $-2.43^{+0.56}_{-0.83}$ & $\cdot \cdot \cdot$ \\
log$_{\rm 10}$(S/H) & $\cdot \cdot \cdot$ & $-3.43^{+1.04}_{-2.12}$ & $\cdot \cdot \cdot$ \\
log$_{\rm 10}$(C/O) & $\cdot \cdot \cdot$ & $-1.33^{+0.83}_{-1.19}$ & $\cdot \cdot \cdot$ \\
log$_{\rm 10}$(N/O) & $\cdot \cdot \cdot$ & $-1.03^{+0.47}_{-0.64}$ & $\cdot \cdot \cdot$ \\
log$_{\rm 10}$(S/O) & $\cdot \cdot \cdot$ & $-1.97^{+0.90}_{-2.04}$ & $\cdot \cdot \cdot$ \\
log$_{\rm 10}$(C/N) & $\cdot \cdot \cdot$ & $-0.39^{+1.08}_{-1.02}$ & $\cdot \cdot \cdot$ \\
$\left[M/H\right]_{\odot}$\tablefootmark{b} & $\cdot \cdot \cdot$ & $74^{+129}_{-53}$ & $\cdot \cdot \cdot$ \\
\hline
\multicolumn{4}{l}{Bayesian evidence} \\
$\ln Z$ & $\cdot \cdot \cdot$ & 451.11 & $\cdot \cdot \cdot$ \\
\hline                                 
\end{tabular*}
\tablefoottext{a}{Strength of evidence is negligible for $0\leq \Delta \ln Z < 1$, positive for $1\leq \Delta \ln Z < 2.5$, strong for $2.5\leq \Delta \ln Z < 5$ and very strong for $\Delta \ln Z \geq 5$.}
\tablefoottext{b}{Metallicity is reported in units of solar metallicity.}
\endgroup
\end{table}

\subsection{Sensitivity of the retrievals to data reduction strategies}

To assess the robustness of the retrieved abundances, we performed retrievals on all spectra obtained using the different reduction strategies. Figure~\ref{fig:mol_constraints} shows the retrieved volume mixing ratios of H$_2$O, NH$_3$, SO$_2$, and CH$_4$ for each reduction. Four of these correspond to alternative reductions based on the fiducial \texttt{Eureka!} pipeline, including the use of Gaussian Processes, a coarser binning scheme of 0.4~$\mu$m, and a reduction excluding the first spectral channel, which may be affected by detector non-linearities such as the brighter--fatter effect \citep{argyriou_brighter-fatter_2023}. We also performed retrievals on the independent reduction using \texttt{JEXORes}. Finally, we compared the results obtained using the \cite{guilluy_ares_2020} spectrum with the one derived from our own HST/WFC3 reduction adopted in the fiducial analysis. All alternative MIRI reductions, including the independent \texttt{JEXORes} reduction, yield consistent molecular abundances. However, the choice of HST reduction has an impact on the retrieved abundances of H$_2$O and NH$_3$. While CH$_4$ and SO$_2$ remain undetected, with similar upper limits across all reductions, the retrieved H$_2$O abundance increases by approximately 0.7 dex when using the \cite{guilluy_ares_2020} spectrum. Although the uncertainties remain overlapping, this highlights the influence of the near-infrared spectrum, which contains the strongest water absorption features. Similarly, the NH$_3$ posterior distribution obtained with our HST reduction exhibits a pronounced low-abundance tail, extending to log$_{10}$(VMR) $\sim -8$ (see ``Custom HST''panel of Fig.~\ref{fig:mol_constraints}). This broadens the posterior distribution and weakens the significance of the ammonia detection from strong to positive, with $\Delta \ln Z$ decreasing from 4.16 to 1.15. A possible explanation for this difference lies in the use of a slightly broader extraction wavelength range in our HST/WFC3 reduction, as well as a smaller spectral binning scheme compared to the reduction made by \citep{guilluy_ares_2020}.

\subsection{Metallicity, mean-molecular weight and elemental ratios}

\begin{figure}
    \centering
    \includegraphics[width=\linewidth]{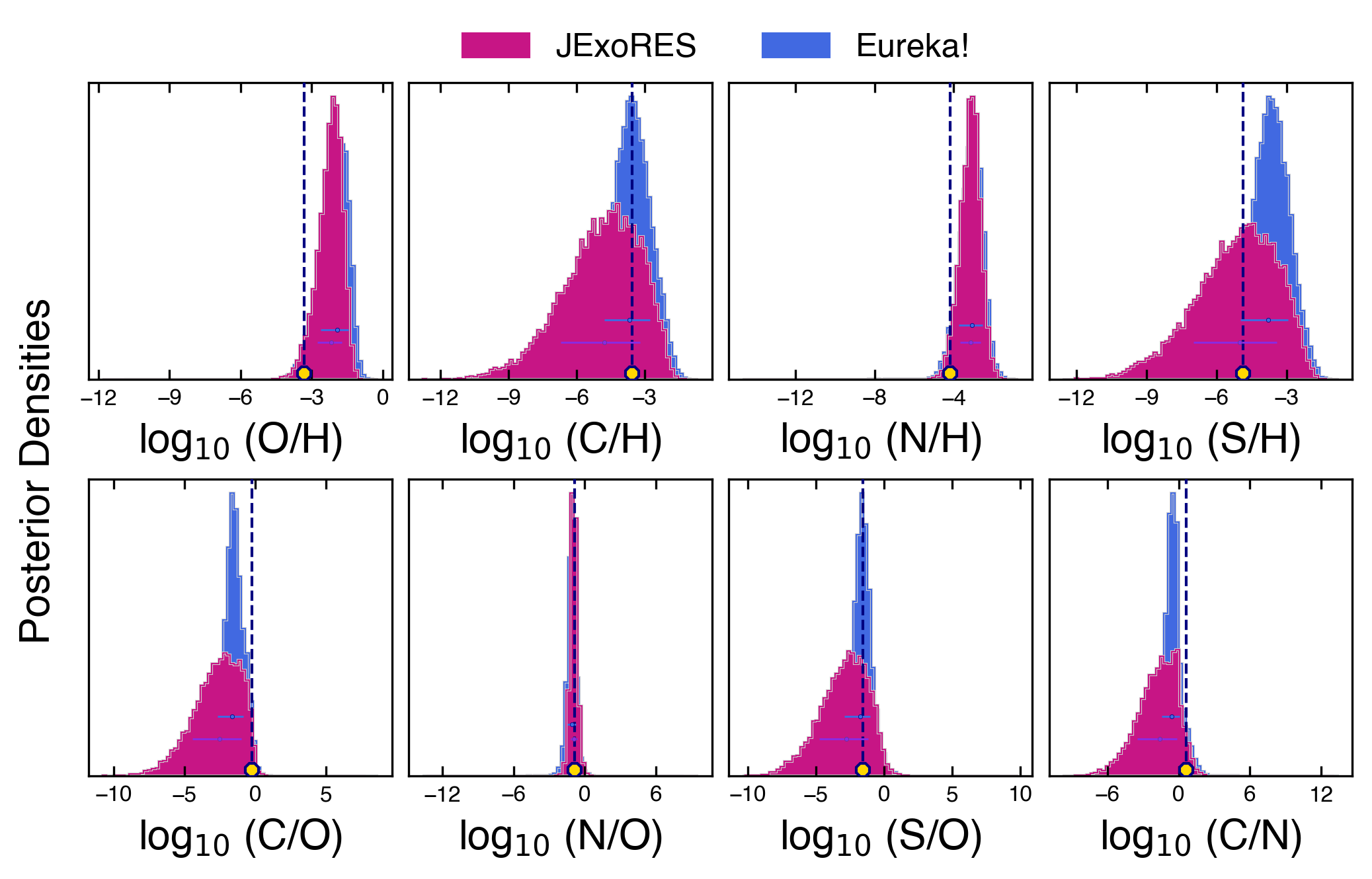}
    \caption{Derived posterior densities of the absolute elemental ratios for the two fiducial reductions of the \texttt{JExoRES} and \texttt{Eureka!} pipelines. Solar values, taken from \cite{asplund_chemical_2021}, are reported with the dashed vertical line for comparison purposes.}
    \label{fig:ratio_constraints}
\end{figure}

We use the samples from the posterior distributions of all our retrievals to determine the constraints on the elemental ratios in the atmosphere of HD~106315~c, which are depicted in Fig.~\ref{fig:ratio_constraints}. With our models providing constraints on the abundances of H$_2$O and NH$_3$, placing upper bounds on the abundance of SO$_2$, CS$_2$, and sometimes very tentatively inferring the presence of OCS, we have constraints on four elements: O, N, S, and C. As no constraints on the elemental ratios of the star HD~106315 are present in the literature, we compared all elemental ratios to solar values. However, we note that the Fe/H ratio of HD~106315 has been consistently found to be sub-solar \citep{crossfield_two_2017, barros_precise_2017, kosiarek_physical_2021}. Our results show super-solar O/H and N/H ratios and a slightly sub-solar nominal value of the N/O ratio, but consistent with the solar value within the uncertainties. Metallicity is calculated as the total abundance of heavy elements relative to hydrogen and expressed in solar units as $ \left[M/H\right]_\odot =
\frac{(C+N+O+S)/H}{8.59\times10^{-4}} $ where $8.59\times10^{-4}$ is the adopted solar heavy-element H-normalised abundance summing contributions from C, N, S and O \citep{asplund_chemical_2021}. With the lack of detections of C-bearing species and only upper constraints on S-bearing species, the clear detection of H$_2$O remains our valuable tracer of atmospheric metallicity. Water abundance scales with the overall oxygen content of the atmosphere and, under many formation scenarios, increases with metallicity. This has been demonstrated in both retrievals and self‑consistent models of exoplanet atmospheres, where higher H$_2$O mixing ratios generally correspond to higher bulk metallicities. We calculated the mean molecular weight (MMW) for each posterior sample by weighting the molecular masses of all species by their retrieved volume mixing ratios, with the remaining atmospheric fraction assigned to H$_2$ and He according to the fixed He/H$2$ ratio. We obtain a MMW $=3.30^{+1.29}{-0.72}$, consistent with an H$_2$/He-dominated atmosphere and comparable to values inferred for other warm Neptune atmospheres \citep{gressier_jwst-tst_2025}.

\section{Forward modelling with \texttt{VULCAN}}

In addition to the retrieval results, we compute a grid of forward models to explore the impact of disequilibrium chemistry on the atmospheric composition of HD~106315~c. For this, we use the \texttt{VULCAN} chemical kinetics code \citep{tsai_vulcan_2017, tsai_comparative_2021}, which solves time-dependent chemical reactions, including photochemistry and vertical mixing, to predict the vertical distribution of molecular abundances under non-equilibrium conditions. We compute a grid of 180 models spanning wide ranges of values of four key parameters: equilibrium temperature $T_{\rm eq}$ (400, 600, 800, 1000, 1200~K), metallicity $\left[\rm M/H\right]_\odot$ (1, 10, 100, 300), carbon-to-oxygen ratio $\rm C/O$ (0.2, 0.8, 1$\times$ Solar, with Solar $\rm C/O = 0.458$), and sulphur-to-hydrogen ratio $\rm S/H$ (0.1, 1, 10$\times$ Solar). These wide ranges are motivated by the uncertainties in the retrieved parameters. Each model is initialised using \texttt{FastChem} \citep{stock_fastchem_2022}\footnote{\url{https://github.com/NewStrangeWorlds/FastChem}} to establish chemical equilibrium abundances, which are then evolved with \texttt{VULCAN} using a fixed vertical mixing coefficient $K_{\rm zz} = 10^{10}$\,cm$^2$\,s$^{-1}$ and a chemical network that includes sulphur species. The vertical mixing coefficient is chosen to be broadly consistent with the value at 1 mbar from the fitting analysis conducted by \cite{moses_chemical_2022}. Our models use the updated network that includes carbon--sulphur chemical coupling, also independently developed by \cite{veillet_inclusion_2026}. The $P- T$ profile is a Guillot Double-Gray temperature profile appropriate for each $T_{\rm eq}$ and metallicity \citep{guillot_radiative_2010}, assuming $\kappa_{\rm th} = 0.2$ and $\kappa_{\rm vis} = 0.02$ cm$^2$ g$^{-1}$. The stellar model adopted is our derived SED UV spectrum for HD~106315 (see Sect.~\ref{sec:stellar_sed} and Appendix~\ref{app:stellar_sed}). This approach allows us to systematically explore how variations in these parameters influence the production and distribution of SO$_2$ and other sulphur-bearing molecules, and to compare these predictions with our retrieval results. 

\section{Discussion}\label{sec:discussion}

\subsection{Water abundance and atmospheric metallicity}

\begin{figure*}
    \centering
    \includegraphics[width=0.5\paperwidth]{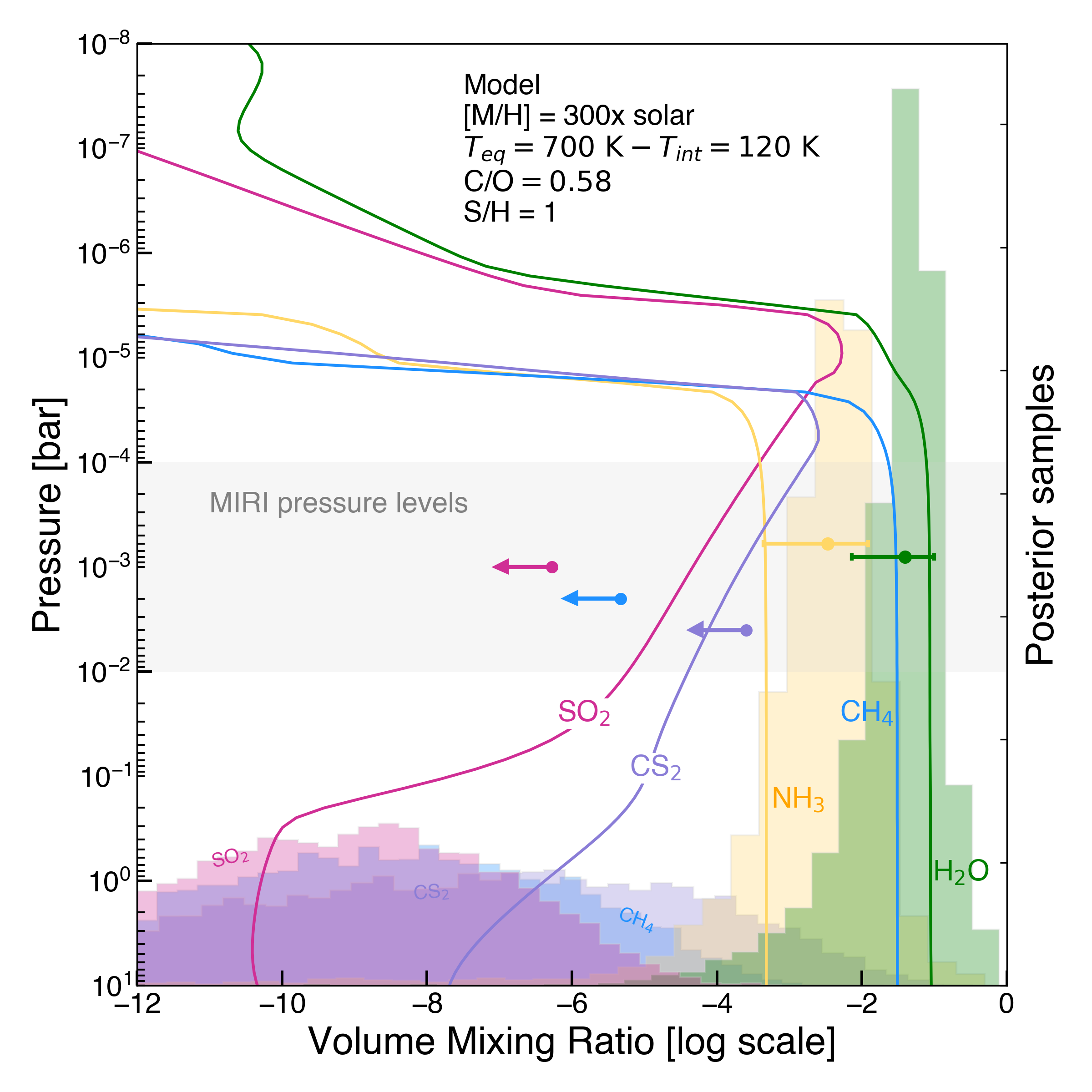}
    \caption{Comparison of results from the \texttt{VULCAN} grid modelling including photochemistry and the atmospheric free-retrieval. Coloured dashed lines show the abundance profiles from the \texttt{VULCAN} grid as a function of pressure, at an equilibrium temperature of 700 K and a metallicity 300$\times$ solar. The histograms are the posterior distributions from the free retrieval. They represent the average abundances between 10$^{-2}$ and 10$^{-4}$ bar. The solid lines are the retrieved median and 1$\sigma$ errors, while the arrows represent retrieved upper limits.}
    \label{fig:model_vs_retrieval}
\end{figure*}

The most robust result from our atmospheric retrievals is the detection of H$_2$O at an abundance of approximately 1 to 10\% by volume. Such a high water abundance places strong constraints on the atmospheric metallicity of HD~106315~c. Fig.~\ref{fig:model_vs_retrieval} compares the retrieved posterior distributions with the predictions from our chemical models. The model providing the best agreement with the retrievals corresponds to an equilibrium temperature of 700 K and a metallicity of 300$\times$ solar. A metallicity of 100$\times$ solar produces broadly similar results, although with a lower H$_2$O abundance. The preferred model assumes a solar C/O ratio of 0.58.

The precise metallicity remains uncertain. Transmission spectroscopy probes the terminator region rather than the global atmosphere, and the retrieved abundances represent average values over the pressure levels contributing to the transmission spectrum. Plus, our model grid samples metallicity at discrete intervals, preventing a more precise determination within the range explored. Nevertheless, the retrieved water abundance provides one of the strongest constraints on the atmospheric composition of HD~106315~c and suggests a highly enriched atmosphere compared to the Solar System giant planets. This result is broadly consistent with the emerging picture that Neptune-mass planets possess atmospheres significantly enriched in heavy elements relative to their host stars \citep[e.g.][]{beatty_sulfur_2024}, which is also consistent with Uranus and Neptune \citep{moses_atmospheric_2020}. However, the inferred enrichment for HD~106315~c lies at the upper end of values currently reported for warm Neptune atmospheres. If confirmed, such a high metallicity would have important implications for both the formation and chemical evolution of the planet. The high water abundance also provides a critical boundary condition for interpreting the atmospheric chemistry. Once the metallicity is constrained by H$_2$O, the abundances of other molecules become highly predictive in the forward models. In particular, the preferred models simultaneously predict detectable abundances of CH$_4$, CS$_2$ and SO$_2$, providing an opportunity to test our understanding of carbon and sulphur chemistry in warm Neptune atmospheres.

\subsection{The missing methane problem and the CH$_4$--CS$_2$ connection}

For the metallicities favoured by the water abundance and equilibrium temperatures between 650 and 800\,K, our disequilibrium chemistry models predict detectable abundances of CH$_4$, CS$_2$, and SO$_2$, within the pressure range probed by MIRI (see Fig.~\ref{fig:model_vs_retrieval}). While the retrieved NH$_3$ and H$_2$O abundances are broadly consistent with these predictions, neither CH$_4$, SO$_2$ nor CS$_2$ are detected.

The absence of methane is particularly intriguing. At the temperatures inferred for HD~106315~c, thermochemical equilibrium generally favours CH$_4$ as the dominant carbon-bearing molecule, especially in atmospheres with enhanced metallicities. Instead, our retrievals place an upper limit on CH$_4$ that lies more than two orders of magnitude below the nominal model prediction. HD~106315~c might therefore join a growing population of exoplanets exhibiting the so-called ``missing methane problem'', where observed CH$_4$ abundances are substantially lower than expected from equilibrium chemistry \citep{yu_unusually_2026}. Similar discrepancies have recently been reported for WASP-107~b \citep{dyrek_so2_2024,sing_warm_2024,welbanks_high_2024}, HAT-P-12~b \citep{crouzet_detection_2025,heinke_information_2026} and V~1298~Tau~b \citep{barat_metal-poor_2025}. 

\begin{figure*}
    \centering
    \includegraphics[width=0.88\paperwidth]{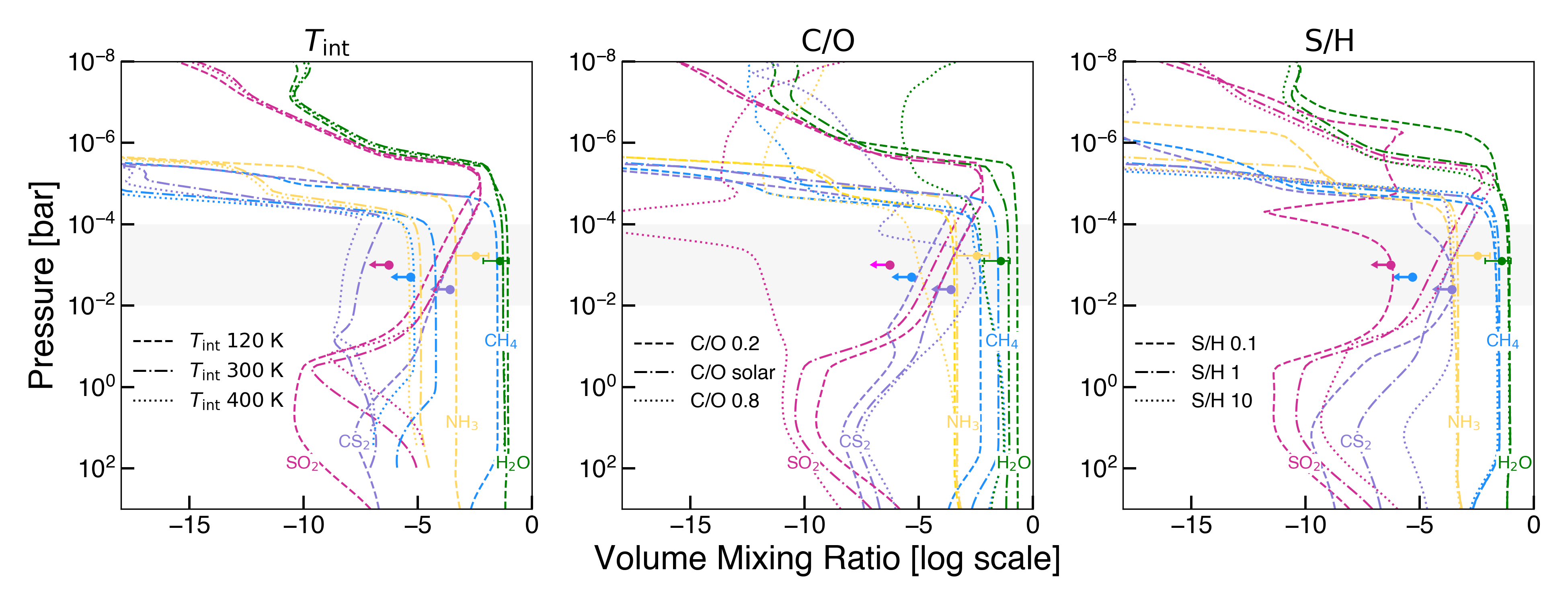}
    \caption{Comparison between the retrieved molecular abundances and disequilibrium chemistry models computed with \texttt{VULCAN}. Coloured curves show the predicted vertical abundance profiles for the different molecules. The shaded region indicates the approximate pressure range probed by the JWST/MIRI transmission spectrum (10$^{-2}-10^{-4}$ bar). The solid lines are the retrieved median and 1$\sigma$ errors, while the arrows represent retrieved upper limits. \textit{Left:} Models with varying intrinsic temperature, where [M/H] = 300$\times$ solar, C/O is solar, and S/H is solar. \textit{Middle:} Models with varying the C/O ratio, assuming $T_{\rm eq}=$700 K, [M/H] = $300\times$ solar, $T_{\rm int}=120$ K, and solar S/H. \textit{Right:} Models with varying S/H ratio, where $T_{\rm eq}=$700 K, [M/H] = 300$\times$ solar, $T_{\rm int}=$120 K, and C/O is solar.}
    \label{fig:model_vs_retrieval_co_sh_tint}
\end{figure*}

The simultaneous non-detection of CS$_2$ may provide an important clue. The recent photochemical model, including an extended C-S chemical network that we use in this work, predicts a strong coupling between CH$_4$ and CS$_2$ abundances in warm, metal-rich atmospheres \citep[e.g. ][]{veillet_inclusion_2026}. In this model we are using, methane acts as a precursor for the production of CS$_2$, such that conditions that suppress CH$_4$ also suppress CS$_2$. Consequently, the absence of CH$_4$ and CS$_2$ in HD~106315~c may not represent two independent discrepancies, but rather the signature of a common physical process. One possibility is that the atmosphere possesses a hotter deep structure than assumed in the nominal models. We therefore computed a set of models increasing the intrinsic temperature, $T_{\rm int}$, to 300 and 400 K from the adopted nominal value of 120 K. Increasing $T_{\rm int}$  modifies the thermal profile at pressures inaccessible to transmission spectroscopy and shifts the methane quench level towards hotter regions of the atmosphere. The left panel of Fig.~\ref{fig:model_vs_retrieval_co_sh_tint} shows the results. Increasing $T_{\rm int}$ leads to a substantial reduction in both CH$_4$ and CS$_2$, bringing the predicted abundances into significantly better agreement with the retrieval constraints. This behaviour naturally arises from the carbon--sulphur coupling \citep{veillet_inclusion_2026}, through which the depletion of methane propagates and suppresses CS$_2$ formation. 
In this test, our forward models predict that the NH$_3$ abundance decreases as the intrinsic temperature increases. This trend is broadly consistent with the retrieved abundance obtained from our fiducial reduction. However, we note that the retrieval based on our independent HST/WFC3 reduction exhibits a pronounced low-abundance tail in the NH$_3$ posterior distribution (see Fig.~\ref{fig:mol_constraints}), allowing substantially lower ammonia abundances. Consequently, the current data do not provide sufficiently robust constraints to distinguish between different intrinsic temperatures based on NH$_3$ alone. We also note that the retrieved $1\sigma$ upper limit on CS$_2$ is higher than that obtained for CH$_4$. Since CS$_2$ is produced from carbon chemistry and its abundance is expected to remain below that of CH$_4$, this result is not physically plausible. It most likely reflects the limited sensitivity of the current MIRI observations to CS$_2$.

For HD~106315~c, the origin of such additional internal heating remains uncertain. Residual heat from formation may contribute to the thermal budget, while ohmic dissipation has been proposed as an additional heat source in highly irradiated atmospheres, although its efficiency in the warm-Neptune regime remains unclear. Tidal heating represents another possibility. The orbital eccentricity of HD~106315~c has been reported as $e = 0.22 \pm 0.15$ \citep{barros_precise_2017}, which could in principle support ongoing tidal dissipation. For example, WASP-107~b is less eccentric \citep{piaulet_wasp-107bs_2021}, but has been found to exhibit substantial tidal heating \citep{wu_evidence_2026}. However, the current uncertainty on the eccentricity of HD~106315~c remains large, and no evidence exists for a resonant configuration or spin-orbit interaction capable of sustaining strong long-term tidal heating \citep{zhou_warm_2018, bourrier_dream_2023,rice_evidence_2023}. Also, HD~106315~c orbits at approximately 0.2 AU and is therefore less strongly irradiated than many of the hot Jupiters for which internal heating mechanisms have been investigated. While an elevated intrinsic temperature provides a plausible explanation for the missing methane and the absence of CS$_2$, it does not resolve the non-detection of SO$_2$. In fact, our models predict that SO$_2$ abundance increases with increasing $T_{\rm int}$, since sulfur partitions into SO$_2$ as CS$_2$ decreases with $T_{\rm int}$. Consequently, although enhanced internal heating may explain the carbon chemistry, an additional mechanism is required to account for the apparent depletion of sulphur-bearing species in the observable atmosphere.

\subsection{Where is the sulphur? Constraints from C/O and S/H}\label{sec:discussion_sulphur}

The non-detection of SO$_2$ is arguably the most surprising result of this study. Sulphur-bearing species have emerged as powerful tracers of atmospheric metallicity and photochemistry in the JWST era \citep{polman_h2_2023, tsai_photochemically_2023}. Sulphur dioxide has now been detected in a growing number of exoplanet atmospheres, including WASP-39~b \citep{alderson_early_2023, rustamkulov_early_2023, powell_sulfur_2024}, WASP-107~b \citep{dyrek_so2_2024, sing_warm_2024, welbanks_high_2024}, GJ~3470~b \citep{beatty_sulfur_2024, madhusudhan_exploring_2025}, HAT-P-26~b \citep{gressier_jwst-tst_2025}, TOI-1130~b \citep{barat_jwst_2026}. In most of these systems, the inferred atmospheric properties are remarkably similar: super-solar metallicities and sub-solar or near-solar C/O ratios, which were predicted for gas giants by \cite{espinoza_metal_2017}. Such conditions favour the efficient photochemical production of SO$_2$ through the oxidation of H$_2$S under UV irradiation from the star \citep{zahnle_atmospheric_2009, tsai_vulcan_2017, polman_h2_2023}.

HD~106315~c appears to satisfy many of the same requirements. The retrieved H$_2$O abundance implies a highly enriched atmosphere, while the host star is an F-type star producing a strong and persistent UV irradiation field. One might therefore expect SO$_2$ to be readily detectable. Instead, our retrievals place stringent upper limits on its abundance. A natural explanation involves variations in the elemental C/O ratio. \cite{polman_h2_2023} showed that SO$_2$ abundance is highly sensitive to C/O, with increasing carbon enrichment suppressing sulphur oxidation pathways. We therefore computed a set of models varying the C/O ratio from the nominal value of 0.58 to 0.2 and to 0.8. The middle panel of Fig.~\ref{fig:model_vs_retrieval_co_sh_tint} shows the resulting volume mixing ratios. Our models reproduce the behaviour described above. Increasing the C/O ratio from the nominal solar value to 0.8 leads to a substantial reduction in SO$_2$ abundance throughout the atmosphere. This occurs because oxygen becomes increasingly locked into carbon-bearing molecules, reducing the amount available for the production of sulphur oxides. Our results show that increasing C/O also affects H$_2$O and NH$_3$. In our models, the same increase in C/O that suppresses SO$_2$ simultaneously decreases the water and ammonia abundances. This creates a tension with the retrievals, which require a large abundance of both molecules, especially H$_2$O. While a modest enhancement above solar C/O cannot be excluded, strongly carbon-rich compositions appear difficult to reconcile with the observed water abundance. The absence of SO$_2$ therefore cannot be explained by an extreme enhancement of C/O.

An alternative possibility is that the atmosphere is intrinsically depleted in sulphur. To investigate this scenario, we explored models with varying S/H ratios presented on the right panel of Fig.~\ref{fig:model_vs_retrieval_co_sh_tint}. Lowering S/H efficiently reduces SO$_2$ abundance while leaving the major oxygen-bearing species largely unchanged. Unlike variations in C/O, sulphur depletion preserves the agreement between the predicted and retrieved H$_2$O and NH$_3$ abundances. Interestingly, the response of CS$_2$ to variations in S/H is significantly weaker than that of SO$_2$. Although both species contain sulphur, their formation pathways are different. SO$_2$ is primarily controlled by photochemical oxidation pathways and therefore scales strongly with the available sulphur inventory. In contrast, CS$_2$ abundance is governed by the coupled carbon--sulphur chemistry that also controls methane abundance. Consequently, changes in S/H do not translate directly into proportional changes in CS$_2$. 

If sulphur is indeed depleted, the origin of this depletion remains unclear. One possibility is that sulphur was incorporated into refractory condensates during planet formation and never efficiently accreted into the atmosphere. Alternatively, sulphur may be sequestered deeper in the atmosphere in species such as H$_2$S that are difficult to observe directly but act as the main reservoir of sulphur. Sulphur depletion has also been proposed in other exoplanet atmospheres and may reflect differences in formation location and migration relative to sulphur-bearing snowlines in the protoplanetary disc \citep{sommerville-thomas_sponchpop_2026}. \cite{sommerville-thomas_sponchpop_2026} indeed suggest that sulphur-poor atmospheres arise when most of the gas is accreted in a region of the disk where sulphur is locked up in solid iron sulfide (FeS), leaving the gas sulphur-poor. For more luminous stars such as HD~106315, the FeS region is larger, which implies that most of the gas envelope was accreted in a region of a few AU, well inside the water snowline. While our observations cannot distinguish between these possibilities, they indicate that the atmospheric sulphur inventory of HD~106315~c differs from that inferred for many previously studied warm Neptune atmospheres.

The stellar irradiation environment provides an additional piece of context. HD~106315 is among the hottest host stars for which detailed sulphur chemistry has been investigated. Compared to systems such as WASP-107~b, HAT-P-26~b, and GJ~3470~b, HD~106315~c receives substantially higher NUV irradiation while experiencing FUV fluxes that remain within the range observed for other warm Neptune atmospheres. Specifically, the NUV flux received by HD~106315~c is approximately 100--200$\times$ larger than that received by WASP-107~b and is comparable to that estimated for WASP-39~b. In contrast, the FUV flux incident on HD~106315~c is similar to that of HAT-P-12~b and HAT-P-11~b (within a factor $\sim 2)$, but remains substantially lower than that of WASP-39~b, which receives nearly two orders of magnitude more FUV irradiation than WASP-107~b. Consequently, the FUV/NUV flux ratio for HD~106315~c is significantly lower than for WASP-107, at $\sim1\%$ of the corresponding value \citep[e.g.][]{france_muscles_2016}.
A note that these comparisons should be taken with caution. The NUV fluxes are generally based on direct observations of the host stars, with the notable exception of WASP-39, for which \cite{tsai_photochemically_2023} adopted HD203244 as a proxy for the NUV spectrum. The FUV estimates are typically more uncertain because they generally rely on stellar proxies and reconstructions. In addition, FUV emission is intrinsically more variable than NUV emission because of stellar activity.
The absence of SO$_2$ despite this intense irradiation suggests that either SO$_2$ gets photochemically destroyed or the availability of sulphur may be more important than the UV flux itself in controlling the observable SO$_2$ abundance. HD~106315~c may represent a chemically distinct regime in which the atmospheric metallicity is constrained by H$_2$O, the deep thermal structure is probed through CH$_4$ and CS$_2$, and the sulphur inventory is traced by the absence of SO$_2$.

To assess whether the unusual irradiation environment of HD~106315~c could explain the observed molecular abundances, we additionally vary the incident FUV flux over a broad range, from $10^{-3}$ to $100\times$ the nominal value. Appendix~\ref{fig:models_low_teq} shows the resulting abundance profiles. Although the upper atmosphere responds noticeably to changes in FUV irradiation, the molecular abundances at the pressures probed by MIRI remain remarkably insensitive to variations of up to an order of magnitude in either direction. Even a reduction of the FUV flux by three orders of magnitude produces essentially unchanged abundances in the MIRI region. A significant change is found for SO$_2$ only for the extreme $100\times$ FUV case. This latter irradiation level is comparable to the FUV environment of substantially more strongly irradiated systems such as WASP-39~b.

These results suggest that the absence of SO$_2$ in HD~106315~c cannot be straightforwardly attributed to an unusually strong FUV field. Given that HD~106315~c occupies an unusual regime of very strong NUV irradiation but comparatively moderate FUV irradiation, NUV-driven chemistry may provide a more relevant avenue for explaining its atmospheric composition. In particular, the combination of strong NUV irradiation and the absence of SO$_2$ motivates further investigation of wavelength-dependent photochemistry rather than interpreting the sulphur abundance solely in terms of the integrated FUV flux. Within the broader irradiation landscape of warm Neptunes, HD~106315~c therefore provides an interesting intermediate case: its FUV environment is not extreme, while its unusually high NUV-to-FUV irradiation balance may distinguish its atmospheric chemistry from that of cooler-host systems such as WASP-107~b.

\subsection{Ammonia as a tracer of cool atmospheric chemistry}

Unlike CH$_4$, CS$_2$ and SO$_2$, ammonia exhibits good agreement between the retrievals and the forward models. Our retrievals favour the presence of NH$_3$ with a median VMR of log${10}$(NH$_3$)$\simeq -3.1$, although the significance of the detection depends on the adopted HST/WFC3 reduction and decreases for the most conservative reductions (see Fig.~\ref{fig:mol_constraints}). The presence of NH$_3$ was previously suggested from HST/WFC3 observations by \cite{guilluy_ares_2020}, although the significance of the detection remained limited. The addition of the JWST/MIRI observations strengthens the evidence for NH$_3$ and places improved constraints on its abundance. 

Our models predict a detectable amount of NH$_3$, which yields quenched NH$_3$ abundances of order $10^{-4}$ over the temperature range favoured by the retrievals ($T_{\rm eq}\sim650$--800 K) and for metallicities above $100\times$ solar. In our models, photodissociation only becomes important below pressures of approximately $10^{-4}$ bar, while the transmission spectrum primarily probes deeper layers. This behaviour contrasts with the results of \cite{ohno_nitrogen_2023_2}, whose photochemical models predict a strong depletion of NH$_3$ at high metallicity. In their calculations, the ammonia abundance collapses at pressures around $10^{-2}$--$10^{-3}$ bar because of photochemical destruction and the increasing dominance of N$_2$ over NH$_3$.

Several factors may explain this apparent discrepancy. First, our adopted Eddy diffusion coefficient ($K_{zz}=10^{10}$ cm$^2$ s$^{-1}$) is one to two orders of magnitude larger than the value used by \cite{ohno_nitrogen_2023_2}, increasing the efficiency of vertical mixing and replenishing NH$_3$ faster than it can be destroyed photochemically \citep[See Figure 3 of ][]{ohno_nitrogen_2023_2}. Second, HD~106315 is an F-type star with a spectral energy distribution that differs significantly from the stellar spectra considered in previous studies. Indeed, the apparent presence of NH$_3$ alongside the non-detection of SO$_2$ and CS$_2$ raises the question of how nitrogen chemistry respond to the specific irradiation environment of HD~106315. The photochemistry of NH$_3$ is particularly sensitive to near-UV photons between approximately 190 and 220 nm, which can penetrate to pressures of $\sim10^{-3}$ bar and efficiently photodissociate ammonia. The balance between these species is therefore expected to depend not only on the total UV flux but on the detailed shape of the stellar UV spectrum.

The absence of detectable HCN may provide an additional clue. In current photochemical models, HCN is produced through reaction pathways involving both NH$_3$ and CH$_4$. Since neither HCN nor CH$_4$ is detected in HD~106315~c, the lack of methane may limit the conversion of NH$_3$ into HCN, allowing ammonia to remain the dominant observable nitrogen-bearing species. Finally, since transmission spectroscopy is most sensitive to the terminator region, we may detect NH$_3$ even if it is not abundant throughout the atmosphere \citep{triantafillides_identification_2026}. Recent JWST observations of WASP-107~b have indeed shown that NH$_3$ may survive preferentially on the cooler morning terminator while being destroyed on the irradiated dayside \citep{murphy_evidence_2024}. A similar day--night asymmetry could contribute to the ammonia abundance inferred for HD~106315~c.

Another possibility is that part of the nitrogen budget is stored in ammonium salts. Recent studies have suggested that NH$_3$ can react with photochemically produced species to form condensates such as ammonium salts in warm exoplanet atmospheres \citep{nakazawa_sulfur_2026}. These condensates may provide an additional reservoir of nitrogen and could alter the observable abundance of NH$_3$. Although our data do not allow us to test this scenario directly, it highlights that the retrieved NH$_3$ abundance may not necessarily trace the total nitrogen content of the atmosphere.

The possible detection of NH$_3$ provides a rare opportunity to investigate the nitrogen chemistry of a warm Neptune atmosphere \citep{fortney_beyond_2020, ohno_nitrogen_2023_1, ohno_nitrogen_2023_2}. Nitrogen is expected to be mainly stored in N$_2$ and NH$_3$, with HCN becoming important when photochemistry is active \citep{lodders_atmospheric_2002, moses_compositional_2013}. Unlike C- and O-bearing species, which can be traced through several molecules, NH$_3$ and HCN are among the few observable tracers of the atmospheric nitrogen content \citep{hobbs_chemical_2019, ohno_nitrogen_2023_1}. Nitrogen abundances are particularly interesting because they may contain information about planet formation. Several studies have suggested that elemental ratios involving nitrogen, such as N/O, C/N, and S/N, can help constrain where a planet formed and how it migrated through the protoplanetary disc \citep{piso_role_2016, cridland_connecting_2020, ohno_jupiters_2021, turrini_tracing_2021, pacetti_chemical_2022}. In particular, the N/O ratio is expected to vary with distance from the star because N- and O-bearing species condense at different temperatures \citep[see Fig.~1 by ][]{ohno_nitrogen_2023_1}. Combining N/O with the more commonly used C/O ratio may therefore provide stronger constraints on formation history than either ratio alone. Using the retrieved NH$_3$ and H$_2$O abundances, we derive a slightly sub-solar nominal N/O ratio, with N/O ranging from approximately $0.0274$ to $0.274$, compared to the solar value of $0.13$. These values lie within the range predicted for the gas phase of protoplanetary disks, where oxygen depletion through condensation enhances the gaseous N/O ratio \citep{ohno_jupiters_2021,
oberg_jupiters_2019}. While the retrieved median N/O is slightly sub-solar, the uncertainties remain large, preventing a strong distinction between gas-dominated and solid-dominated accretion histories.

In general, ammonia detections in exoplanetary atmospheres remain comparatively rare relative to molecules such as H$_2$O, CO$_2$, and CH$_4$. The most robust detections have been obtained in the atmospheres of directly imaged giant planets and substellar objects, where lower temperatures favour the stability of NH$_3$. For example, \cite{malin_first_2025} reported the first unambiguous detection of NH$_3$ in the atmosphere of a directly imaged giant exoplanet, a result later confirmed and refined by \cite{matthews_second_2026}. Ammonia isotopologues have also been detected in brown dwarf atmospheres, including the first detection of $^{15}$NH$_3$ reported by \cite{barrado_15nh3_2023}. In transiting exoplanets, evidence for NH$_3$ has only recently begun to emerge. A first detection of nitrogen chemistry was reported by \cite{macdonald_hd_2017} in the atmosphere of the hot Jupiter HD~209458~b. Tentative detections were reported in the atmosphere of the Saturn-mass planet WASP-107~b \citep{welbanks_high_2024} and the warm giant WASP-80~b by \cite{wiser_precise_2025}, followed by a more robust identification by \cite{triantafillides_identification_2026}. In our case, NH$_3$ would represent one of the few tentative detections of ammonia in a transiting exoplanet atmosphere and one of the first in a Neptune-mass planet.

\subsection{SO$_2$ and CS$_2$ shorelines around solar-type stars}\label{sec:shoreline}

Recent work by \cite{crossfield_mapping_2025} introduced the concept of an SO$_2$ ``shoreline'', separating regions of equilibrium temperature and metallicity where SO$_2$ is expected to be detectable from those where it is not. This framework was developed primarily using the warm-Saturn mass planet HAT-P-26~b orbiting a K-type star and has proven useful for interpreting the growing sample of SO$_2$ detections obtained with JWST. An important question is whether this shoreline remains valid for planets orbiting hotter stars. HD~106315 is an F-type star with a substantially different UV spectrum than the K/M-dwarf hosts for which SO$_2$ was detected, considered by \cite{crossfield_mapping_2025}. Since SO$_2$ production is fundamentally driven by photochemistry, differences in stellar spectral energy distributions may alter both the location and shape of the shoreline.

\begin{figure*}
    \centering
    \includegraphics[width=\linewidth]{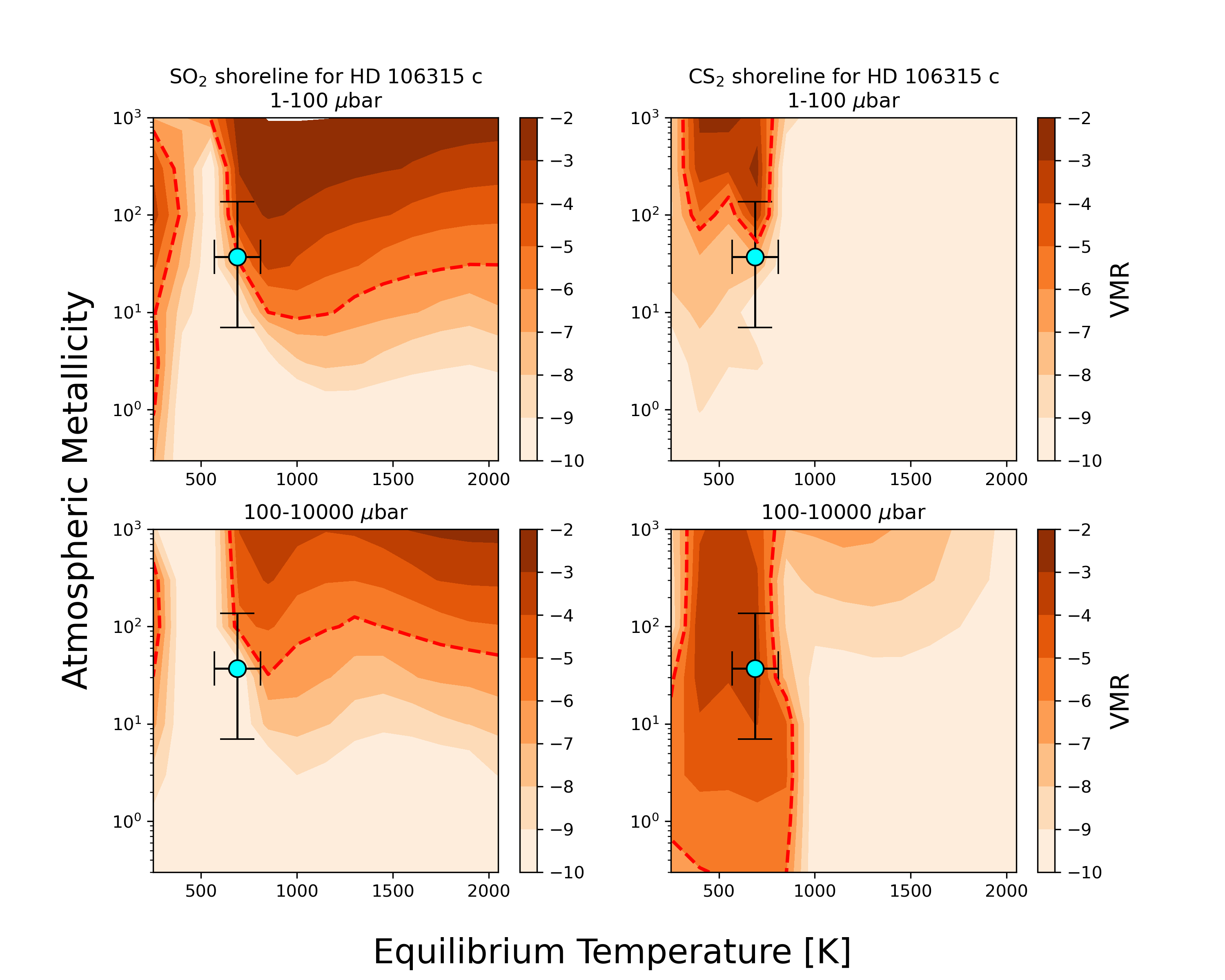}
    \caption{Temperature--metallicity maps showing the predicted pressure-weighted abundances of SO$_2$ and CS$_2$ for a Solar-type host star. Colours indicate the VMR SO$_2$ and CS$_2$, while the dashed red contour corresponds to a VMR of 1 ppm. The cyan marker and associated error bars indicate the atmospheric properties inferred for HD~106315~c from our retrievals. \textit{Upper-left:} SO$_2$ abundance averaged over the photochemically active region of the atmosphere (1--100 $\mu$bar) following the prescription of \cite{crossfield_mapping_2025}. \textit{Lower-left:} SO$_2$ abundance averaged over deeper atmospheric layers (100--10 000 $\mu$bar),  which corresponds to the approximate pressure range probed by the JWST/MIRI transmission spectrum. \textit{Upper-right:} CS$_2$ abundance averaged over the photochemically active region of the atmosphere (1--100 $\mu$bar). \textit{Lower-right:} CS$_2$ abundance averaged over the deeper layers (100--10 000 $\mu$bar).}
    \label{fig:shorelines}
\end{figure*}

To investigate this possibility, we computed a grid of \texttt{VULCAN} models spanning equilibrium temperatures from 250 to 2050 K and metallicities from solar to $3000\times$ solar. Unlike previous work, we explored two different methods for estimating observable SO$_2$ abundances. The first follows the approach of \citet{crossfield_mapping_2025} but uses pressure-weighted average abundances across the photochemically active region between 1 and 100 $\mu$bar. The second pressure-weighted averages are over deeper atmospheric layers extending from 100 to 10 000 $\mu$bar, motivated by the knowledge that MIRI transmission spectroscopy probes pressures closer to the millibar level. Fig.~\ref{fig:shorelines} shows the computed shorelines. Our result show that the shorelines differ significantly from those obtained for the K-type star atmosphere. Indeed, we identify a low-temperature regime in which SO$_2$ abundance increases again at equilibrium temperatures below approximately 500 K. This behaviour is absent from the original shoreline and appears to be either a consequence of the different photochemical environment produced by an F-type host star or the use of the updated chemical network including the carbon--sulphur coupling \citep{veillet_inclusion_2026}.

To understand the origin of this behaviour, we examine the vertical abundance profiles of SO$_2$ and CS$_2$ across a range of temperatures. These results are presented Fig.~\ref{fig:models_low_teq}. At intermediate temperatures between approximately 400 and 500 K, carbon--sulphur chemistry efficiently produces CS$_2$, which becomes the dominant sulphur-bearing species. However, as temperatures decrease further, the kinetic pathways responsible for CS$_2$ formation become increasingly inefficient. The collapse of the CS$_2$ abundance at 250 K allows sulphur to be redistributed towards alternative reservoirs, including SO$_2$. In this regime, the increase in SO$_2$ abundance is therefore not driven by enhanced photochemistry, but rather by the suppression of a competing carbon--sulphur pathway. This result suggests that SO$_2$ and CS$_2$ should be considered jointly when interpreting sulphur chemistry in exoplanet atmospheres. While SO$_2$ traces photochemical oxidation pathways, CS$_2$ traces the efficiency of carbon--sulphur coupling. Together, they provide a more complete view of sulphur partitioning than either molecule alone. More broadly, our results demonstrate that the SO$_2$ shoreline is not universal. The location of the shoreline depends not only on atmospheric metallicity and temperature, but also on the spectral type of the host star and the pressure levels sampled by the observations.

\subsection{Are clouds muting the atmospheric features?}\label{sect:clouds}

An alternative explanation for the apparent absence of several molecular species is the presence of aerosols that reduce the amplitude of spectral features in transmission. Clouds and hazes are commonly invoked to explain muted transmission spectra in exoplanetary atmospheres and have been inferred in numerous systems observed with HST and JWST. We explore several aerosol prescriptions in our retrieval analysis, including a grey cloud deck, a combined cloud and haze model, and silicate clouds capable of producing opacity features in the MIRI wavelength range. Indeed, silicate clouds are known to display mid-infrared features, especially the Si-O stretching feature at $10\ \mathrm{\mu m}$. For this particular retrieval setup, we focus on the aerosol species MgSiO$_3$ and use the Mie-scaterring slab model presented in \cite{mullens_implementation_2024}. This model is parameterized by the mean particle size $r_m$ in $\mu$m, with a log-uniform prior between -3 and 1, the cloud-top pressure $P_{\rm top, slab}$ in bar, with a log-uniform prior from -8 to 2, the width of the cloud in log-pressure space, $D \log P$, with a uniform prior from 0 to 10, and finally the constant log mixing ratio $\rm log_{10}(VMR)_{\rm MgSiO_3}$ of MgSiO$_3$ in the cloud, with a uniform prior between -30 and -1. The aerosol extinction cross-section is taken from the precomputed database available in \texttt{POSEIDON}\footnote{\url{https://zenodo.org/records/15711943}}.
Table~\ref{table:clouds} shows the log-evidence for each model as well as the log-evidence difference with the no-cloud model, taken as our null hypothesis.  

\begin{table}[h!]
\caption{Retrieved log-evidence values for different cloud and haze parametrisations and comparison to the ``no-clouds'' null hypothesis.}  
\label{table:clouds} 
\begingroup
\renewcommand{\arraystretch}{1.2}
\centering                   
\begin{tabular*}{\columnwidth}{@{\extracolsep{\fill}}lcc}
\hline\hline              
Model &  $\ln Z$ & $\Delta \ln Z$\tablefootmark{a} \\ 
No clouds & 451.39 & $\cdot \cdot \cdot$ \\
Clouds and hazes & 451.23 & -0.16 \\
Cloud deck & 451.11 & -0.28 \\
Silicate clouds (MgSiO$_3$) & 449.79 & -1.16 \\
\hline   
\end{tabular*}
\tablefoottext{a}{Strength of evidence is negligible for $0\leq \Delta \ln Z < 1$, positive for $1\leq \Delta \ln Z < 2.5$, strong for $2.5\leq \Delta \ln Z < 5$ and very strong for $\Delta \ln Z \geq 5$.}
\endgroup
\end{table}

While these models provide acceptable fits to the data, none is statistically preferred over the cloud-free solution. The silicate cloud model is positively disfavoured compared to the no-cloud model. The absence of a strong aerosol signature does not imply that clouds are entirely absent from the atmosphere. Thin cloud decks located below the pressures probed by transmission spectroscopy may still be present without leaving a measurable imprint on the spectrum. Similarly, aerosol populations with particle size distributions different from those considered here may remain undetected. We also examined the impact of the different prescriptions on the retrieved molecular abundances. The constraints on CH$_4$ and SO$_2$ remain broadly consistent across the cloud-free and cloudy models, with variations of up to $\sim0.9$ dex in the retrieved upper limits. Thus, including clouds does not significantly alter our conclusions regarding the non-detection of these species. The lack of a significant Bayes factor should not be interpreted as evidence for the absence of clouds. The data do not statistically distinguish between the cloudy and cloud-free scenarios. Clouds at pressures probed by transmission spectroscopy therefore remain compatible with the observations.

\subsection{HD~106315~c in the context of warm Neptune atmospheres}

\begin{figure}
    \centering
    \includegraphics[width=\linewidth]{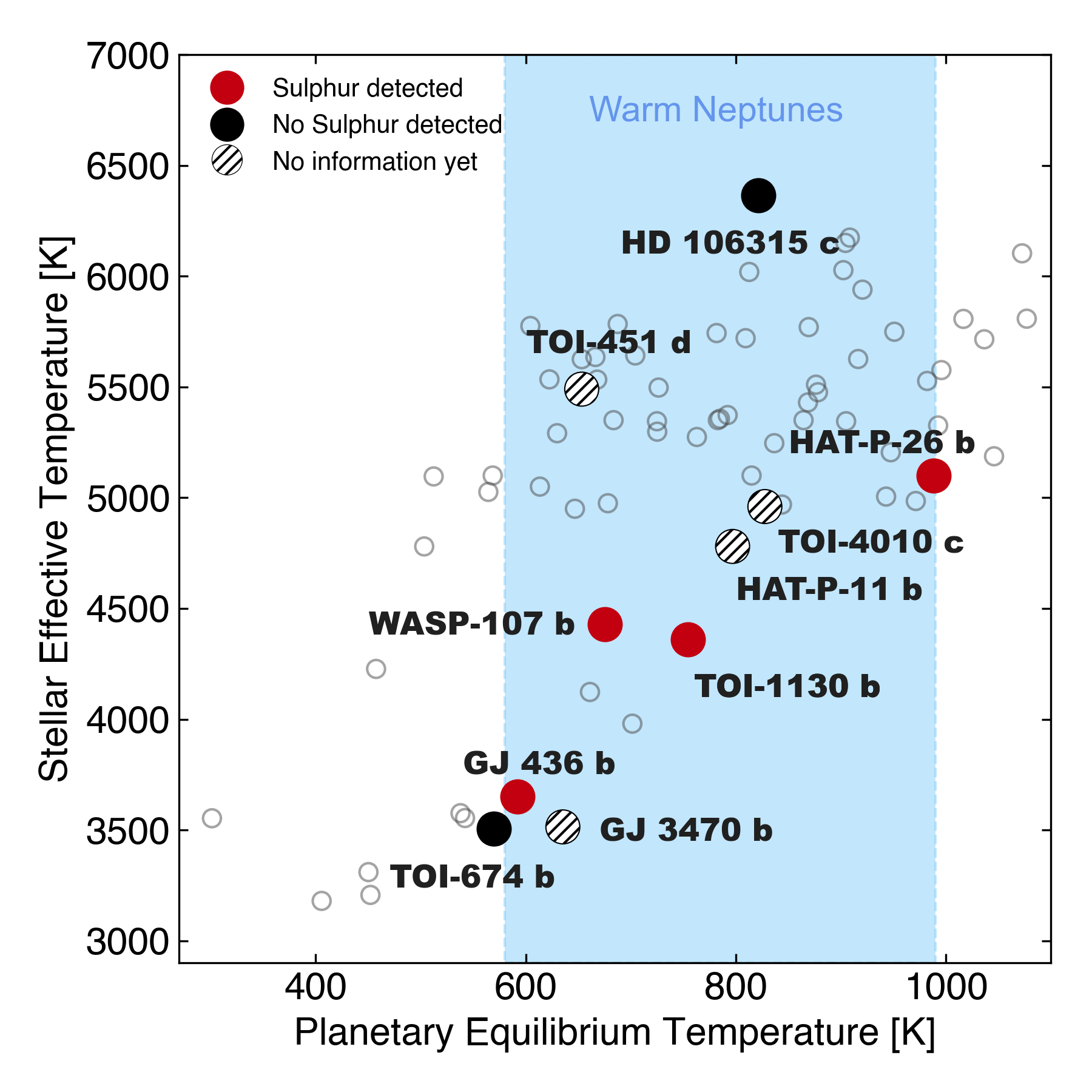}
    \caption{Stellar effective temperature--planetary equilibrium temperature distribution of the warm Neptune population. The blue shaded region marks the $T_{\rm eq}$ range adopted for warm Neptunes in this work. The large cricles denote targets with JWST observations either completed or scheduled, with red and black being published sulphur detections and non-detections respectively. Hatching marks are targets for which no published JWST atmospheric analysis is yet available.}
    \label{fig:population}
\end{figure}

The growing sample of warm Neptune atmospheres observed with JWST has revealed an unexpectedly rich chemical diversity \citep{gressier_jwst-tst_2025, dyrek_so2_2024, sing_warm_2024, welbanks_high_2024, beatty_sulfur_2024, madhusudhan_exploring_2025, barat_jwst_2026}. Molecular species detected in this population now include H$_2$O, CH$_4$, CO$_2$, CO, tentative NH$_3$, SO$_2$ and several sulphur-bearing compounds. Fig.~\ref{fig:population} shows the distribution of the warm Neptune population in the stellar effective temperature--planetary equilibrium temperature plane, highlighting the current JWST observational coverage and the targets for which sulphur has been detected. The sample is defined by $R_{\rm p} > 3\;R_\oplus$, $T_{\rm eq} < 1000$ K, $10 < M_{\rm p}/M_\oplus < 35$, and $P < 25$ days. Most of these planets orbit M- and K-type host stars, where favourable planet-to-star radius ratios and deep transits enable high signal-to-noise atmospheric characterisation. HD~106315~c represents the first detailed atmospheric studies of a warm Neptune orbiting an F-type star. In contrast to these previously observed systems and as discussed thoroughly in previous sections of this paper, we find robust evidence for H$_2$O and tentative evidence for NH$_3$ but no evidence for CH$_4$, CS$_2$ or SO$_2$. This difference may reflect genuine chemical diversity within the warm Neptune population. Alternatively, it may partly arise from observational limitations. Compared to planets orbiting later-type stars, HD~106315~c transits a substantially larger host star and therefore its transmission signal is smaller. The atmospheric scale-height signal is therefore intrinsically weaker, making molecular detections more challenging even with JWST. As the sample of warm Neptune atmospheres continues to grow, comparisons across stellar spectral types will become increasingly important for disentangling the roles of irradiation, metallicity and elemental composition in shaping atmospheric chemistry.

\subsection{The need for additional insights from near-infrared spectroscopy}\label{sect:nir_data}

Our predictions through modelling reveal a substantial number of possibilities regarding the chemistry at play in the atmosphere of HD~106315~c, and therefore the need to break degeneracies arising from the data. In addition, the absence of both SO$_2$ and CS$_2$ suggests that the sulphur inventory of the atmosphere is not yet fully accounted for. A major limitation of the current dataset is the lack of direct constraints on the carbon budget. The abundance of CH$_4$ is only weakly constrained through an upper limit, while CO and CO$_2$ remain undetected. 


Near-infrared observations with JWST/NIRISS and NIRSpec would directly address this limitation by probing the strong absorption bands of H$_2$O, NH$_3$, CH$_4$, CO$_2$, and CO between 0.6 and 5.2~$\mu$m. In particular, the NIRSpec/G395H mode \citep[$R\sim2700$, ][]{birkmann_near-infrared_2022} covers the fundamental bands of CO$_2$ near 4.3~$\mu$m and CH$_4$ near 3.3~$\mu$m, while NIRISS/SOSS \citep{albert_near_2023} would provide additional constraints on H$_2$O, NH$_3$, and atmospheric scattering processes and clouds.
Additional constraints on the carbon inventory would help to distinguish between the scenarios described above. In particular, measurements of CH$_4$, CO$_2$, and CO would provide direct constraints on the atmospheric C/O ratio and on the degree of chemical disequilibrium. Combined with the existing detections of H$_2$O and NH$_3$, such measurements would allow a more complete assessment of the relative abundances of carbon-, nitrogen-, and sulphur-bearing species. More broadly, the combination of detectable NH$_3$ and apparently depleted sulphur species suggests that nitrogen and sulphur chemistry may respond differently to the irradiation environment of F-type stars, highlighting the need for a more comprehensive understanding of atmospheric chemistry across different stellar spectral types. Finally, combined with the existing MIRI spectrum, these observations would yield the first panchromatic transmission spectrum (0.6--12~$\mu$m) of a warm Neptune orbiting an F-type star.



\section{Conclusion}

We have presented the JWST/MIRI transmission spectrum of the warm Neptune HD~106315~c, obtained as part of the JWST-SUPER program (PID 2950, PI: R. Waters) and combined with archival HST/WFC3 observations (PID 15333, PIs: I. Crossfield \& L. Kreidberg) and a contemporaneous reconstruction of the stellar high-energy irradiation environment. HD~106315~c is one of the first Neptune-mass exoplanets orbiting an F-type star to be studied in detail with JWST, providing a valuable opportunity to investigate atmospheric chemistry in an irradiation regime that differs substantially from the M- and K-type hosts that dominate current atmospheric samples.

Our atmospheric retrievals consistently reveal a water-rich atmosphere and provide tentative evidence for ammonia, with retrieved abundances of $\log_{10}({\rm H_2O})=-1.40^{+0.40}_{-0.74}$ and $\log_{10}({\rm NH_3})=-2.47^{+0.57}_{-0.89}$, together with a terminator temperature of $747^{+150}_{-155}$ K. In contrast, we find no evidence for CH$_4$, SO$_2$, CS$_2$, CO$_2$, CO, HCN, H$_2$S, tentative evidence for OCS, and derive stringent upper limits on their abundances. The retrieval results are robust against different data reduction strategies and independent retrieval frameworks, indicating that the inferred atmospheric composition is not driven by methodological choices. The detection of abundant H$_2$O implies a strongly metal-enriched atmosphere, favoured by our disequilibrium chemistry calculations.

The simultaneous presence of H$_2$O and NH$_3$, combined with the absence of CH$_4$, is particularly intriguing. At the temperatures probed by the observations, methane is generally expected to remain an important carbon reservoir, while ammonia is often predicted to be depleted by photochemistry. HD~106315~c is therefore likely to join the subset of exoplanets exhibiting the ``missing-methane problem'' \citep{yu_unusually_2026}. Our \texttt{VULCAN} disequilibrium chemistry models show that the retrieved NH$_3$ abundance can be reproduced for highly enriched atmospheres, especially when efficient vertical mixing replenishes ammonia from deeper atmospheric layers. The apparent survival of NH$_3$ may also reflect the specific UV spectral shape of the F-type host star \citep{ohno_nitrogen_2023_2}, day--night chemical asymmetries at the terminator \citep{murphy_evidence_2024}, or the presence of additional nitrogen reservoirs such as ammonium salts \citep{nakazawa_sulfur_2026}. Regardless of the underlying mechanism, HD~106315~c joins a very small group of transiting exoplanets showing evidence for atmospheric ammonia and may represent one of the first such cases in a Neptune-mass planet.

One of the primary goals of the SUPER program is to understand the origin and observability of sulphur-bearing molecules. In this context, the absence of both SO$_2$ and CS$_2$ is one of the most important outcomes of this study. Recent JWST observations have revealed SO$_2$ in several warm exoplanets, including WASP-39~b, HAT-P-26~b, GJ~3470~b, and WASP-107~b, establishing sulphur photochemistry as a major tracer of atmospheric irradiation and metallicity \citep{alderson_early_2023, rustamkulov_early_2023, powell_sulfur_2024, dyrek_so2_2024, sing_warm_2024, welbanks_high_2024, beatty_sulfur_2024, madhusudhan_exploring_2025, gressier_jwst-tst_2025, barat_jwst_2026}. HD~106315~c differs from these planets in a significant way: despite orbiting a hot F-type star and receiving substantial UV irradiation, it exhibits no detectable sulphur-bearing species. Our photochemical models predict that a highly enriched atmosphere should produce observable SO$_2$, CS$_2$, or both. Increasing the intrinsic temperature suppresses methane and CS$_2$ through carbon--sulphur coupling \citep{veillet_inclusion_2026} and helps reconcile the CH$_4$ and CS$_2$ upper limits, but it does not fully explain the absence of SO$_2$. Instead, our model suggest that the observations are more readily explained if the atmospheric sulphur abundance is lower than expected from a scaled-solar composition, or if HD~106315~c lies near a transition between distinct sulphur-chemistry regimes.

To investigate this possibility, we compute a new grid of disequilibrium chemistry models appropriate for a Solar-type irradiation environment and revisited the recently proposed SO$_2$ ``shoreline'' framework \citep{crossfield_mapping_2025}. We find that the location of the SO$_2$ shoreline depends not only on equilibrium temperature and metallicity, but also on the stellar spectral energy distribution and the atmospheric pressure levels probed by the observations. In particular, the inclusion of carbon--sulphur coupling introduces a low-temperature regime in which CS$_2$ formation becomes inefficient and sulphur is redistributed into alternative reservoirs. Our results therefore suggest that SO$_2$ and CS$_2$ should be interpreted jointly, as complementary tracers of sulphur partitioning rather than independent indicators of photochemistry. More generally, the shorelines derived for K- and M-type host stars cannot necessarily be extrapolated to planets orbiting hotter stars.

We also investigate whether high-altitude aerosols could explain the absence of molecular features. Retrievals including grey clouds, hazes, and spectrally active silicate clouds all provide acceptable fits to the data, but none is statistically preferred over the cloud-free solution. While clouds may still exist below the pressures probed by transmission spectroscopy, our observations provide no evidence that aerosols alone are responsible for the non-detections of CH$_4$, SO$_2$, or CS$_2$. 

Taken together, these results reveal an atmospheric composition that differs from those of several previously characterised warm Neptunes \citep{gressier_jwst-tst_2025, dyrek_so2_2024, sing_warm_2024, welbanks_high_2024, beatty_sulfur_2024, madhusudhan_exploring_2025, barat_jwst_2026}. While a fraction of Neptune-mass planets observed with JWST exhibit signatures of sulphur photochemistry, HD~106315~c instead shows abundant H$_2$O, tentative NH$_3$, and no detectable sulphur-bearing molecules. This demonstrates that UV irradiation alone is not sufficient to guarantee the presence of observable SO$_2$, and highlights the complex interplay between atmospheric metallicity, elemental abundance ratios, internal heat transport, vertical mixing, and host-star spectral type. 

Finally, our results illustrate the importance of extending atmospheric studies beyond planets orbiting cool stars. Most currently characterised warm Neptunes orbit M- and K-type hosts, whereas HD~106315~c probes a fundamentally different photochemical environment around an F-type star. Future JWST observations with NIRISS and NIRSpec will be crucial for directly constraining the planet's carbon inventory through measurements of CH$_4$, CO$_2$, and CO, thereby testing whether the apparent methane depletion and sulphur deficiency share a common origin. Combined with the existing MIRI spectrum, such observations would provide the first panchromatic transmission spectrum of a warm Neptune orbiting an F-type star and enable a more complete assessment of the coupled carbon, nitrogen, and sulphur chemistry operating in this emerging class of exoplanets.

\begin{acknowledgements}  
We thank Jeroen Bouwman and Migo M\"uller and O. D. Renes for their significant work and contribution to the preparation of the observations and data reduction. We thank Ian Crossfield for insightful discussions on the SO$_2$ shoreline.
This research is based on observations made with the NASA/ESA Hubble Space Telescope obtained from the Space Telescope Science Institute, which is operated by the Association of Universities for Research in Astronomy, Inc., under NASA contract NAS5–26555. These observations are associated with program 15333 (PIs: I. Crossfield \& L. Kreidberg). This work is based on observations made with the NASA/ESA/CSA James Webb Space Telescope. The data were obtained from the Mikulski Archive for Space Telescopes at the Space Telescope Science Institute, which is operated by the Association of Universities for Research in Astronomy, Inc., under NASA contract NAS5-03127 for JWST. These observations are associated with program 2950 (PI: R. Waters).
This research is based on observations obtained with XMM-Newton, an ESA science mission with instruments and contributions directly funded by ESA Member States and NASA. We acknowledge the use of public data from the \textit{Swift} data archive.
B. Edwards acknowledges the use the Dutch national e-infrastructure with the support of the SURF Cooperative using grant no. EINF-15133.
J.P.P. thanks the Swift project team, and especially Kim Page (UK Swift Science Data Centre at University of Leicester) for their advice in planning and analysis of the Swift observations,
and Simon Rosen (ESA XMM-Newton Science Operations Centre ESAC) for advice on the characteristics of the OM instrument, and especially the UV-grism and its data.
J.P.P.\ acknowledges financial support from the UK Science and Technology Facilities Council, and the UK Space Agency, during part of this work.
S-.M. T. is supported by the National Science and Technology Council (grants 114-2112-M-001-065-MY3) and an Academia Sinica Career Development Award (AS-CDA-115-M03). S-.M. T. acknowledges the access to the Kawas cluster provided by Academia Sinica Institute of Astronomy and Astrophysics (ASIAA).
For the purpose of open access, the authors have applied a Creative Commons Attribution (CC BY) licence to the Author Accepted Manuscript version arising from this submission.\\
{\bf Facilities used:} 
NASA Astrophysics Data System;
CDS Simbad \& Vizier database systems;
NASA Exoplanet Archive;
ExoplANETS-A host-stars database;
Neil Gehrels Swift Observatory;
Swift Data Archive;
XMM-Newton Observatory;
XMM Data Archive;
James Webb Space Telescope Observatory;
Hubble Space Telescope Observatory;
SVO Filter Profile Service `Carlos Rodrigo', funded by MCIN/AEI/10.13039/501100011033/ through grant PID2023-146210NB-I00 \\ 
{\bf Software used:} 
HEASoft package (including the \textit{Swift} UVOT tools), XMM Scientific Analysis System (SAS), TOPCAT/STILTS, WebPIMMS, DS9, ARIADNE \citep{Vines2022}, \texttt{XSPEC} \citep{arnaud1996}, \texttt{Python} \citep{Python3}, \texttt{numpy} \citep{numpy,harris2020array}, \texttt{matplotlib} \citep{matplotlib}, \texttt{scipy} \citep{2020SciPy-NMeth}, \texttt{corner} \citep{corner}, \texttt{astropy} \citep{astropy:2013, astropy:2018, astropy:2022}, \texttt{MultiNest} \citep{buchner_johannesbuchnermultinest_2021},\texttt{dynesty} \citep{speagle_dynesty_2020}, \texttt{celerite} \citep{celerite1}, \texttt{UltraNest} \citep{buchner_ultranest_2021}, \texttt{JExoRES} \citep{holmberg_exoplanet_2023}, \texttt{CASCADe} \citep{lahuis_horizon_2020,carone_indications_2021}, \texttt{jwst} \citep{bushouse_jwst_2025}, \texttt{Eureka!} \citep{bell_eureka_2022}, \texttt{batman} \citep{kreidberg_batman_2015},\texttt{POSEIDON} \citep{macdonald_hd_2017, macdonald_poseidon_2023}, \texttt{TauREx} \citep{al-refaie_taurex_2021}, \citep{tsai_vulcan_2017, tsai_comparative_2021}.

\end{acknowledgements}

\bibliographystyle{bibtex/aa}
\bibliography{references.bib}   

\clearpage
\begin{appendix}
\onecolumn
\section{Data analysis} \label{section:reduction comparison}
\subsection{Non-linearity correction} \label{section:custom linearity}
\begin{figure*}[h!]
    \centering
    {\includegraphics[width=0.8\paperwidth] {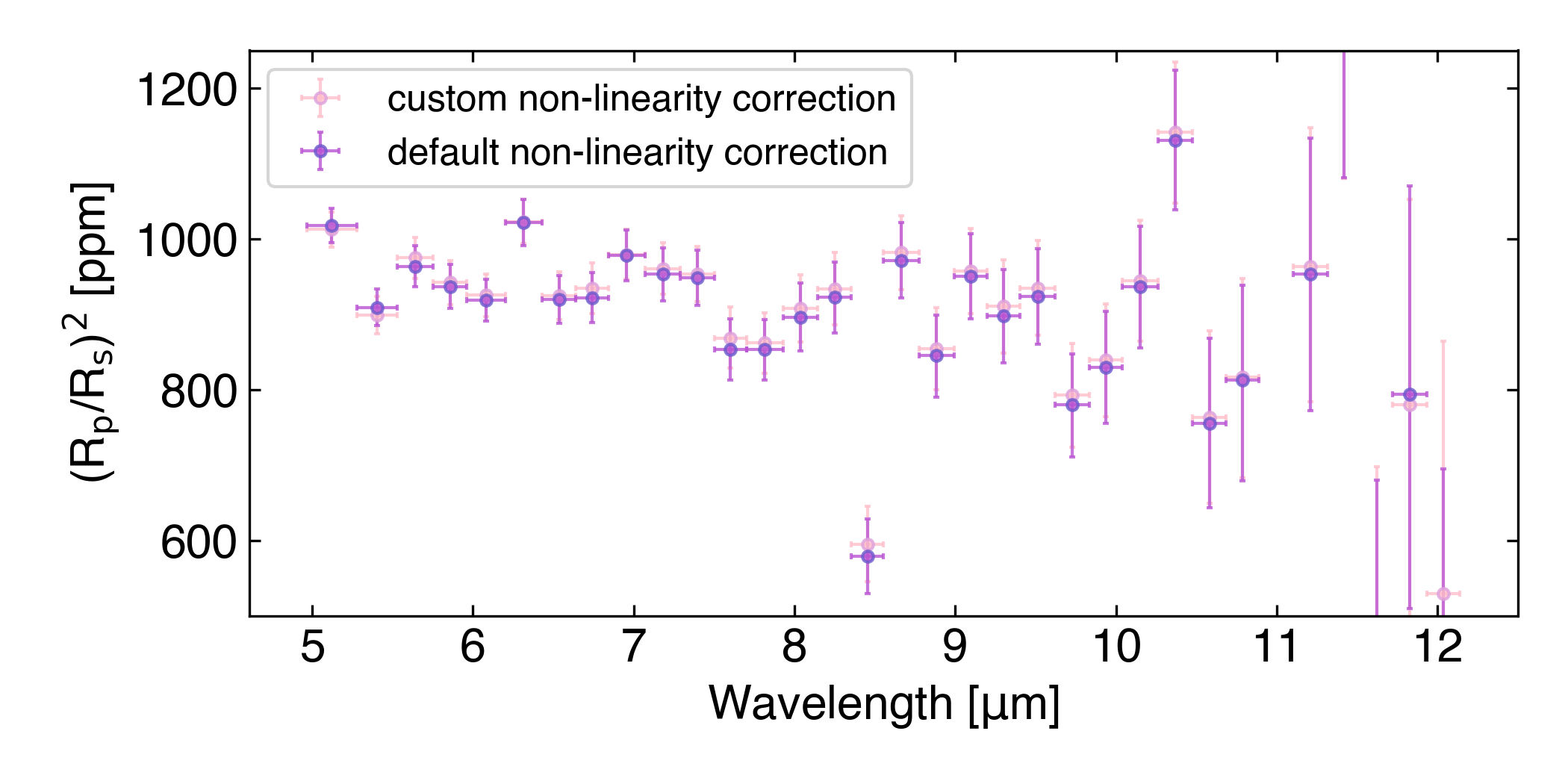}}
     \caption{MIRI transmission spectrum of HD~106315~c using the default (from the \texttt{jwst} pipeline) and the custom non-linearity corrections. These spectra do not use GPs and errorbars may be underestimated.}
      \label{fig:spectrum_non_linearity}
\end{figure*}

Non-linearities in the MIRI detector have been shown to be problematic for several reasons. For example, the beginning of the ramp is affected by multiple reset-related non-linearities, such as the Reset-Switch Charge Decay \citep[RSCD,][]{ressler_reset_2025} and the Brighter-Fatter Effect \citep[BFE,][]{argyriou_brighter-fatter_2023}. The final frame is affected by the last-frame effect, and, more generally, the signal accumulation is non-linear, with a signal-dependent decrease in the ramp slope.
Applying the default linearity correction of the \texttt{jwst} pipeline \citep{bushouse_jwst_2025} in Stage~1 as well as flagging the first four frames possibly affected by the RSCD effect substantially improves the linearity of the ramp. However, a residual slope remains when computing pairwise differences between ramps, indicating that the default correction is not yet optimal. As non-linearity effects may substantially impact the derived transit depth, we derive an alternative linearity correction based on the data itself. We fit the following parametric model to the detector ramps
\begin{equation}
\begin{aligned}
\mathrm{S}_{\mathrm{ij}}(t)=a_{i j, 0}+\tau_{i k, 1} \cdot a_{i j, 1} \cdot\left(1-\mathrm{e}^{\frac{-t}{\tau_{i j, 1}}}\right)-\tau_{i j, 2} \cdot a_{i j, 2} \cdot \mathrm{e}^{\frac{-t}{\tau_{i j, 2}}} & \\
& i \in\{0, \ldots, 415\}, \quad j \in\{0, \ldots, 72\}, \quad 0 \leq t \leq \mathrm{T}_{\mathrm{int}}
\end{aligned}
\end{equation}

where $t$ is the time between 0 and the duration of a single integration $\mathrm{T}_{\mathrm int}$. The first term represents the debiasing effect. The second term models the RSCD effects. We fit this model to the detector ramps after applying the reset pipeline step, for all integrations. Using the fitted estimate of the linearised signal, we derive a custom non-linearity correction \citep{dyrek_so2_2024} used as a replacement calibration file in the linearity pipeline step. This procedure consequently improves the linearity of the ramps while not affecting the resulting spectra. The comparison between the extracted spectra with the default non-linearity correction step of the \texttt{jwst} pipeline and the custom non-linearity correction step is shown in Fig.~\ref{fig:spectrum_non_linearity}.

\begin{figure*}[h!]
    \centering
    {\includegraphics[width=0.8\paperwidth] {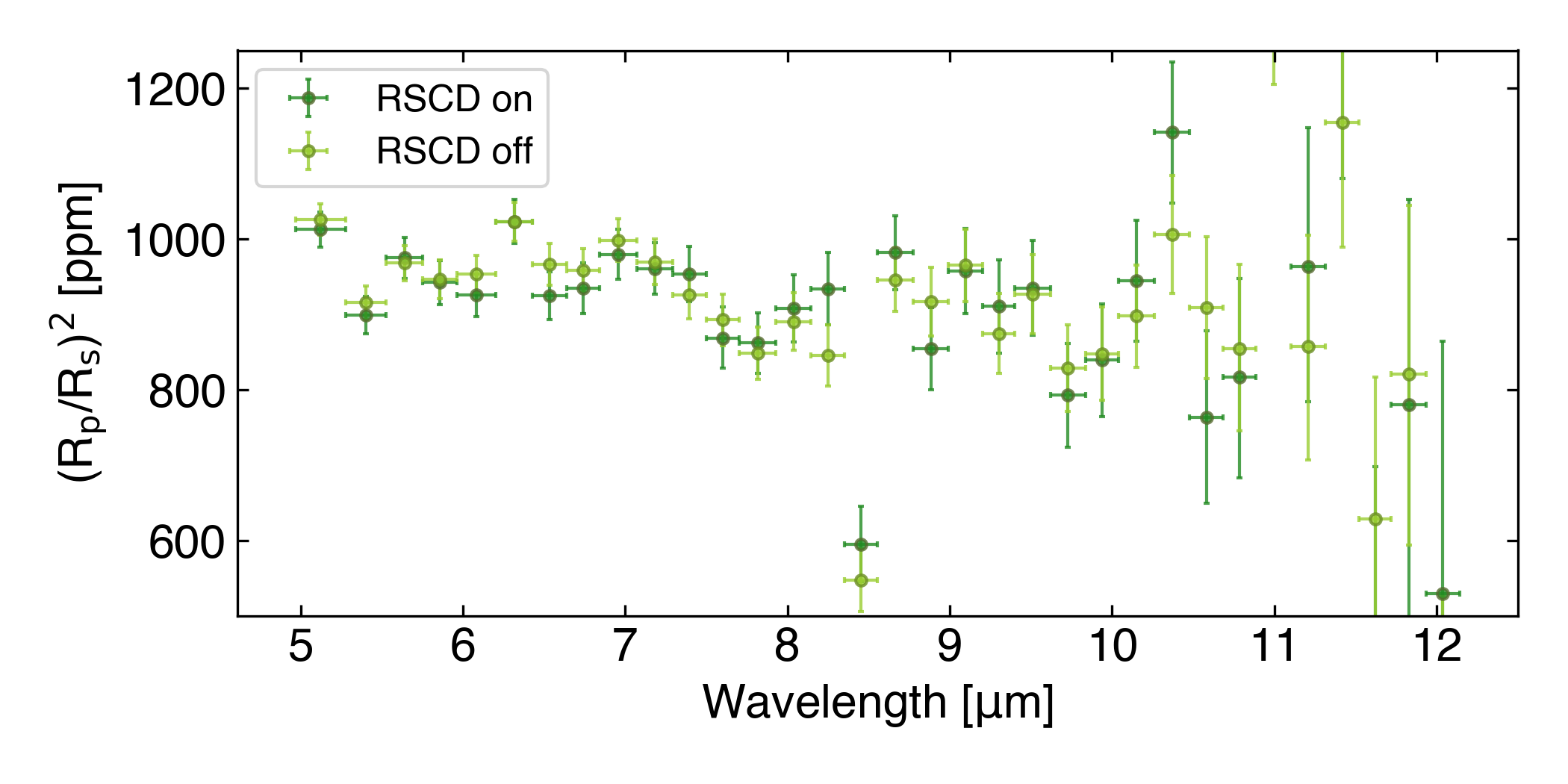}}
     \caption{MIRI transmission spectrum of HD~106315~c switching on and off the RSCD step in the \texttt{jwst} pipeline. These spectra do not use GPs and error bars may be underestimated.}
      \label{fig:spectrum_rscd}
\end{figure*}

We also tested leaving the first four frames in the datasets by switching off the RSCD step. The resulting spectra are consistent with each other within the error bars. Fig.~\ref{fig:spectrum_rscd} shows the comparison between the two reductions done with the \texttt{jwst} pipeline with and without activating the RSCD step.

\subsection{Cosmic-ray hit at 8.5 $\mu$m} \label{section:cosmic_hit}

\begin{figure*}[h!]
    \centering
    {\includegraphics[width=0.8\paperwidth] {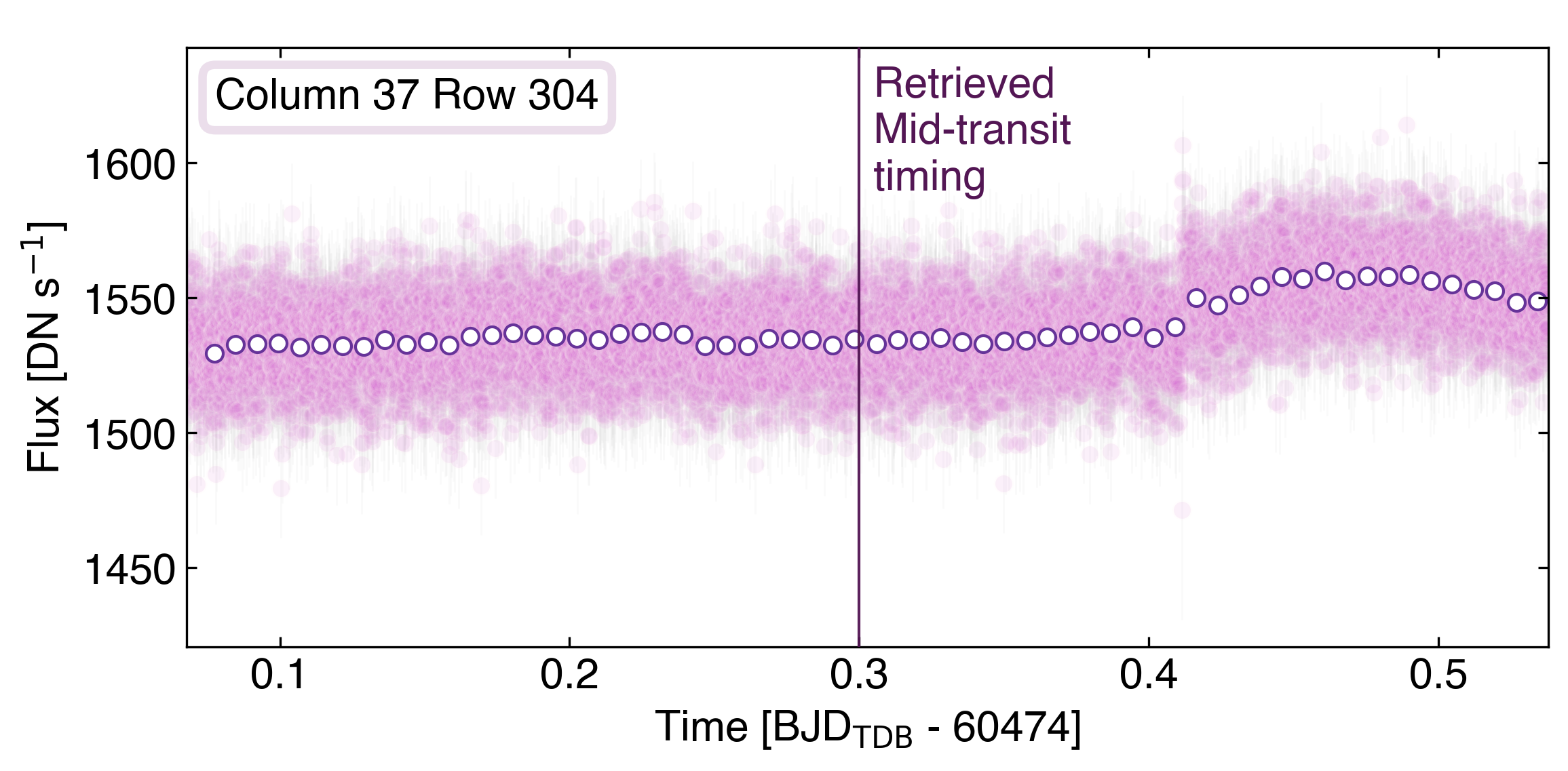}}
     \caption{MIRI time series of the pixel located at row 304 and column 37 of HD~106315~c included in the wavelength bin at ~8.5 $\mu$m. The pixel does not recover from a cosmic hit at $\sim$60474.4 BJD$_{\rm TDB}$.}
    \label{fig:cosmic_hit_pixel}
\end{figure*}

As discussed in Sect.~\ref{section:robustness}, the spectral point near 8.5~$\mu$m in the MIRI data appears significantly lower than the neighbouring spectral points (see Fig.~\ref{fig:best_fit_baseline_spec}). To investigate the origin of this feature, we examined the individual detector pixels contributing to this wavelength bin. Fig.~\ref{fig:cosmic_hit_pixel} shows the time series of one representative pixel (row 304, column 37). The flux remains stable throughout the first part of the observation but exhibits a sudden increase shortly after the transit event, at approximately BJD$_{\rm TDB}=60474.4$ days. The timing and amplitude of this discontinuity are consistent with a cosmic-ray hit or a similar detector-level event. Unlike typical cosmic-ray signatures, the affected pixel does not recover from the hit. Instead, the flux remains persistently higher, producing a step-like change in the baseline. Such behaviour is not accounted for by the transit model and can bias the inferred transit depth for wavelength bins containing the affected pixels. In particular, this increase in flux causes the fitted transit depth to be underestimated, leading to the anomalously low spectral point observed near 8.5~$\mu$m. The disappearance of this feature when adopting a larger 0.4~$\mu$m binning scheme further supports the interpretation that it originates from a localised detector effect rather than an atmospheric signal. Nevertheless, because the affected pixels remain included in the wider wavelength bins, we cannot exclude the possibility of some residual influence on neighbouring spectral points. 

\subsection{Spectral binning} \label{section:spectral_binning}

\begin{figure*}[h!]
    \centering
    {\includegraphics[width=0.8\paperwidth] {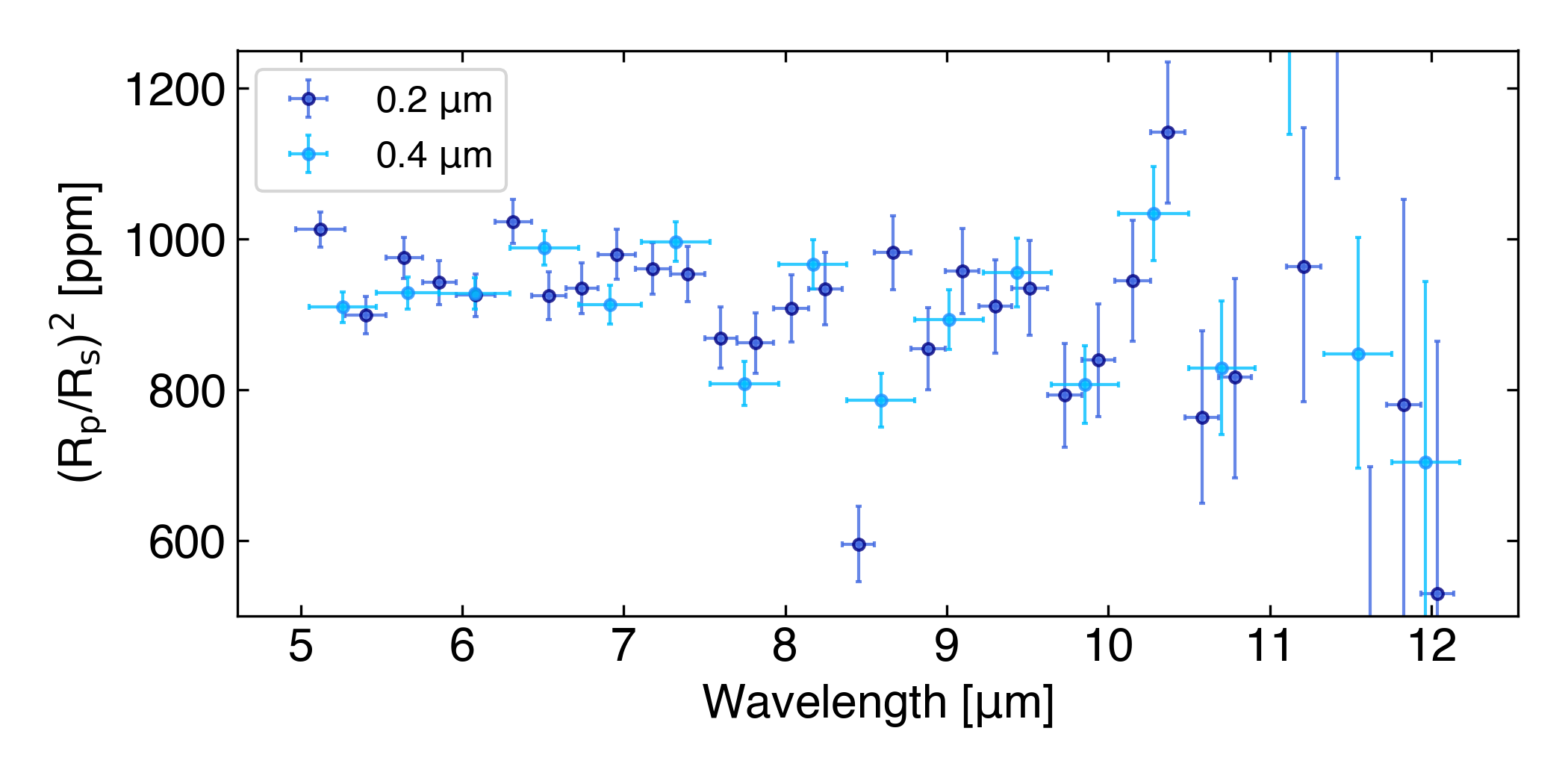}}
     \caption{MIRI transmission spectrum of HD~106315~c using different bin widths. These spectra do not use GPs and errorbars may be underestimated.}
      \label{fig:spectrum_binsize}
\end{figure*}

To assess the robustness of the extracted transmission spectrum with respect to the choice of spectral binning, we repeated the analysis using a larger wavelength binning scheme. Increasing the bin width reduces the impact of photon noise and potential wavelength-to-wavelength correlations even though there is a cost in terms of spectral resolution. Such tests are commonly performed to verify that detected spectral features are not driven by individual wavelength channels or residual systematics affecting a small number of detector rows.
Fig.~\ref{fig:spectrum_binsize} compares the transmission spectra obtained using the nominal (0.2 $\mu$m) and larger (0.4 $\mu$m) binning schemes. The two spectra are consistent within their uncertainties across the full wavelength range, indicating that the inferred spectral features are not sensitive to the adopted bin size. As expected, the larger bins yield smaller uncertainties thanks to the increased signal-to-noise ratio. This agreement demonstrates that the spectral morphology is robust and is not an artefact of the particular wavelength sampling chosen for the nominal analysis.
We also note that the low spectral point near 8.5~$\mu$m observed in the 0.2~$\mu$m spectrum is not apparent in the 0.4~$\mu$m binned spectrum. This behaviour is consistent with the feature being caused by a localised detector effect whose influence is diluted when combining a larger number of wavelength channels. However, because the affected pixels still contribute to the wider bin, we cannot completely exclude a residual impact on the neighbouring spectral points. The overall agreement between the two binning schemes nevertheless demonstrates that the spectral morphology is robust and is not driven by our choice of wavelength sampling.

\subsection{Time-correlated noise} \label{section:time-correlated noise}

\begin{figure*}[h!]
\centering
\includegraphics[width=0.4\paperwidth]{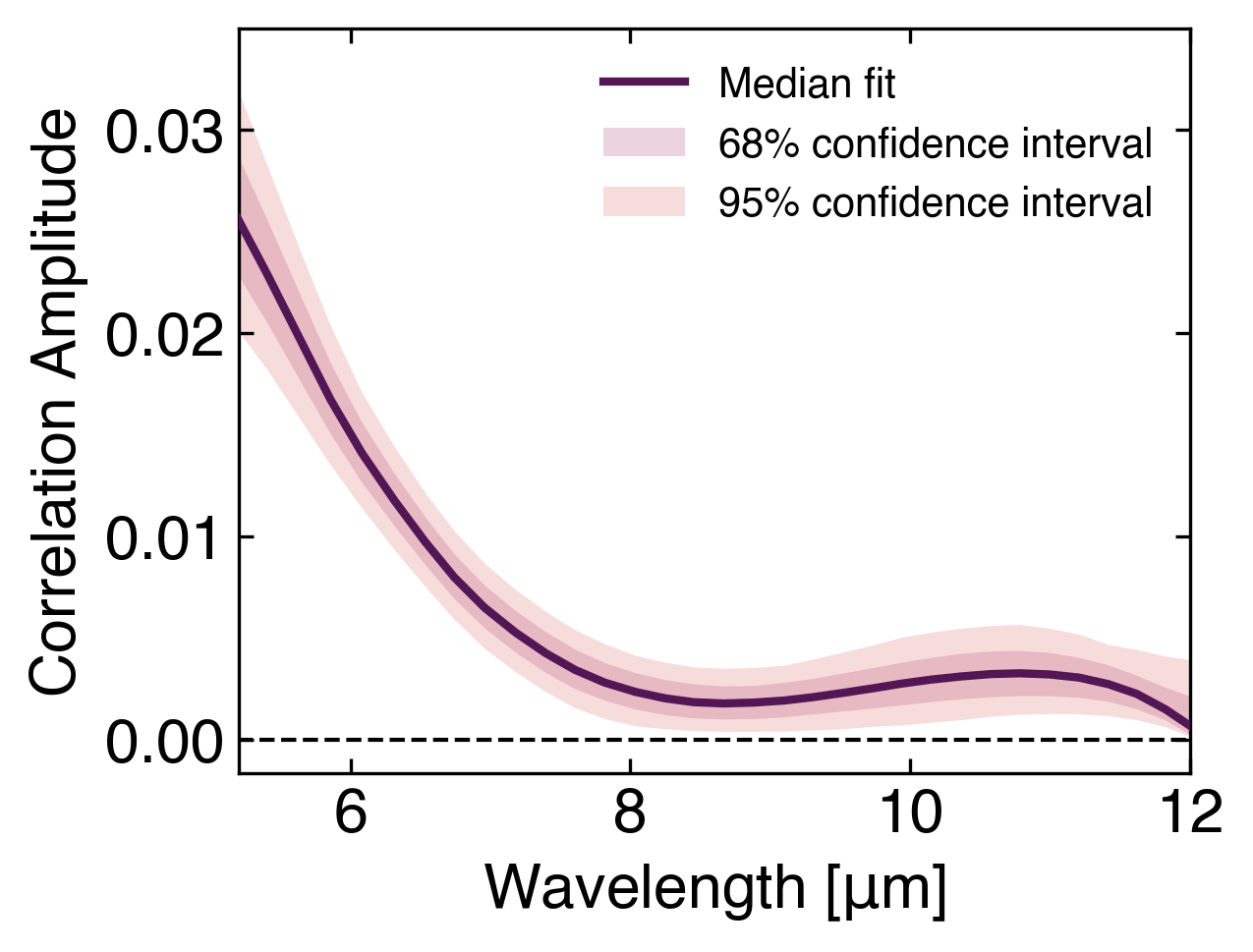}
\includegraphics[width=0.4\paperwidth]{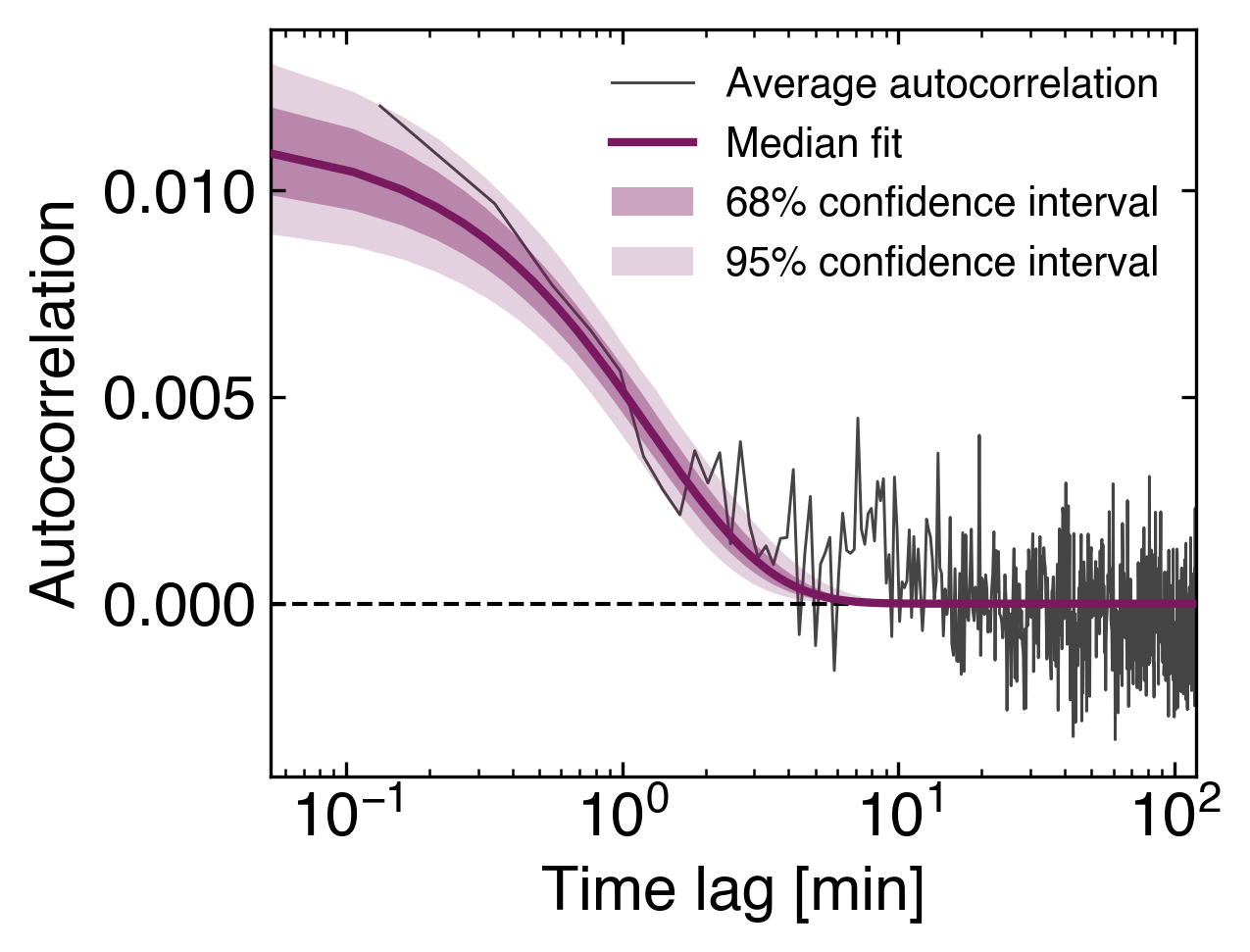}
  \caption{Left panel: time-dependent noise correlation amplitude. Right panel: time-dependent noise autocorrelation function. The median fits are
shown in solid line, while the 68\% and 95\% confidence intervals are shown in lighter shades.}
     \label{fig:correlated_noise}
\end{figure*}

Time-correlated noise is a well-known challenge in exoplanet transit spectroscopy because it can bias inferred transit depths and lead to underestimated uncertainties when neglected \citep{cubillos_correlated-noise_2017}. Such effects have been reported in various observations \citep[e.g.][]{holmberg_exoplanet_2023, sarkar_exoplanet_2024}. For JWST/MIRI observations in particular, detector systematics remain an active area of investigation due to the instrument's low operating temperature (~6 K) leading to a complex recovery behaviour of individual pixels following charge trapping. As a result, Gaussian Process (GP) models have become a common tool within the exoplanet community to account for residual correlated noise while marginalising over its impact on the retrieved planetary parameters \citep{fortune_how_2024}. To characterise the structure of the time-correlated noise in our observations, we employ the method proposed by \citet{Madhusudhan2025}. We model the correlated noise using a GP with an exponential kernel and fit the kernel hyperparameters using the light-curve residuals. The GP amplitude is allowed to vary with wavelength, as described by a low-order polynomial. Parameter estimation is performed using the \texttt{celerite} package \citep{celerite1} with the nested-sampling algorithm implemented in \texttt{UltraNest} \citep{buchner_ultranest_2021}. The resulting wavelength-dependent correlation amplitudes and the average autocorrelation function are shown in Fig.~\ref{fig:correlated_noise}. For more details, see \citet{Madhusudhan2025}.

The strength of the correlated noise varies significantly with wavelength. The inferred correlation amplitude is largest at short wavelengths, reaching values of approximately 2--3\% near 5~$\mu$m, before decreasing steadily toward longer wavelengths. Beyond $\sim$8~$\mu$m, the correlation amplitudes are substantially smaller, although a weak secondary enhancement may be present around 10--11~$\mu$m. This wavelength dependency suggests that the dominant source of correlated noise is not purely temporal but is coupled to detector properties that vary across the spectral trace.
We also find clear evidence for correlated noise on short timescales. The average autocorrelation function has the largest values at short time lags and decreases rapidly with increasing lag, becoming consistent with zero after approximately 3--5 minutes. This behaviour indicates that the dominant correlated-noise component operates on timescales comparable to a small number of integrations. To account for these effects in the final transmission-spectrum extraction, we incorporate this wavelength-dependent GP noise model's amplitude and time scale values into both the \texttt{Eureka!} and \texttt{JExoRES} analyses. This approach allows us to propagate the impact of correlated noise into the inferred transit depths and their uncertainties. 

\section{Free-chemistry atmospheric retrieval}\label{section:appendix_retrievals}

\begin{figure*}[h!]
    \centering
    {\includegraphics[width=0.9\paperwidth] {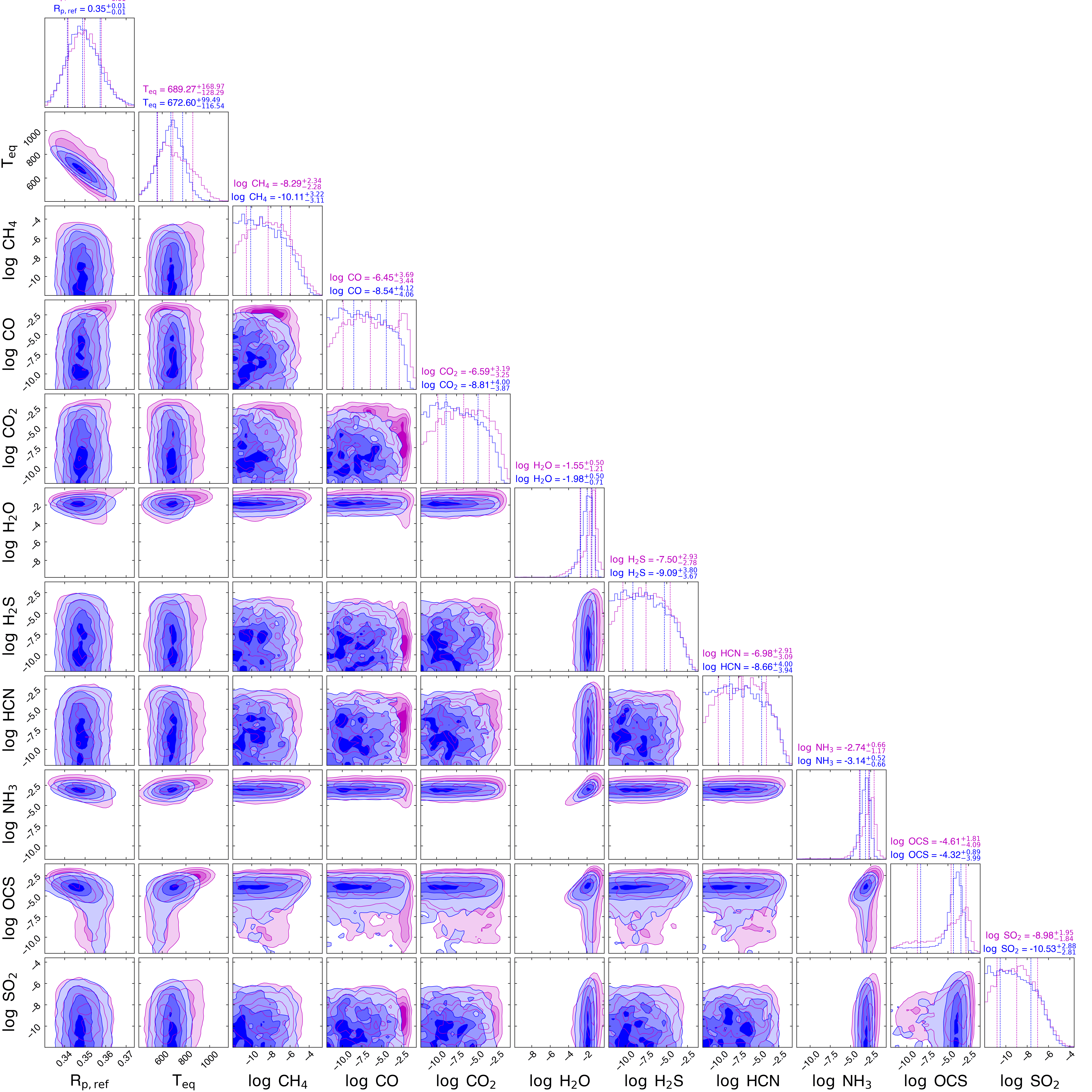}}
     \caption{Posterior distributions from the free retrievals of the MIRI and WFC3 spectra of HD~106315~c, respectively using the \texttt{Eureka!} and our own HST/WFC3 reductions. The blue posterior distribution corresponds to the \texttt{Taurex} retrieval, while the purple one is from \texttt{POSEIDON}.}
    \label{fig:corner_plot}
\end{figure*}

As indicated in Sect.~\ref{section:retrievals}, we performed independent free-chemistry retrievals using the \texttt{TauREx} and \texttt{POSEIDON} retrieval frameworks to validate our atmospheric analysis. Both retrievals were carried out using the same input dataset, consisting of the HST/WFC3 transmission spectrum obtained from our independent reduction and the JWST/MIRI transmission spectrum extracted with the \texttt{Eureka!} pipeline. Fig.~\ref{fig:corner_plot} compares the posterior distributions obtained with the two retrieval codes. For clarity, only parameters common to both frameworks are shown. In particular, CS$_2$ is not included in this comparison because the corresponding opacity database is currently unavailable in \texttt{TauREx}. Despite differences in the underlying implementations, sampling strategies, and forward-model architectures of the two codes, the retrieved posterior distributions are in excellent agreement. The median values and credible intervals overlap for all shared parameters, and both frameworks recover the same parameter degeneracies and correlations. This comparison therefore serves as a valuable sanity check, demonstrating that our atmospheric inferences are not driven by specific features in the retrieval codes implementation or the underlying models.

\begin{figure*}
    \centering
    \includegraphics[width=0.5\linewidth]{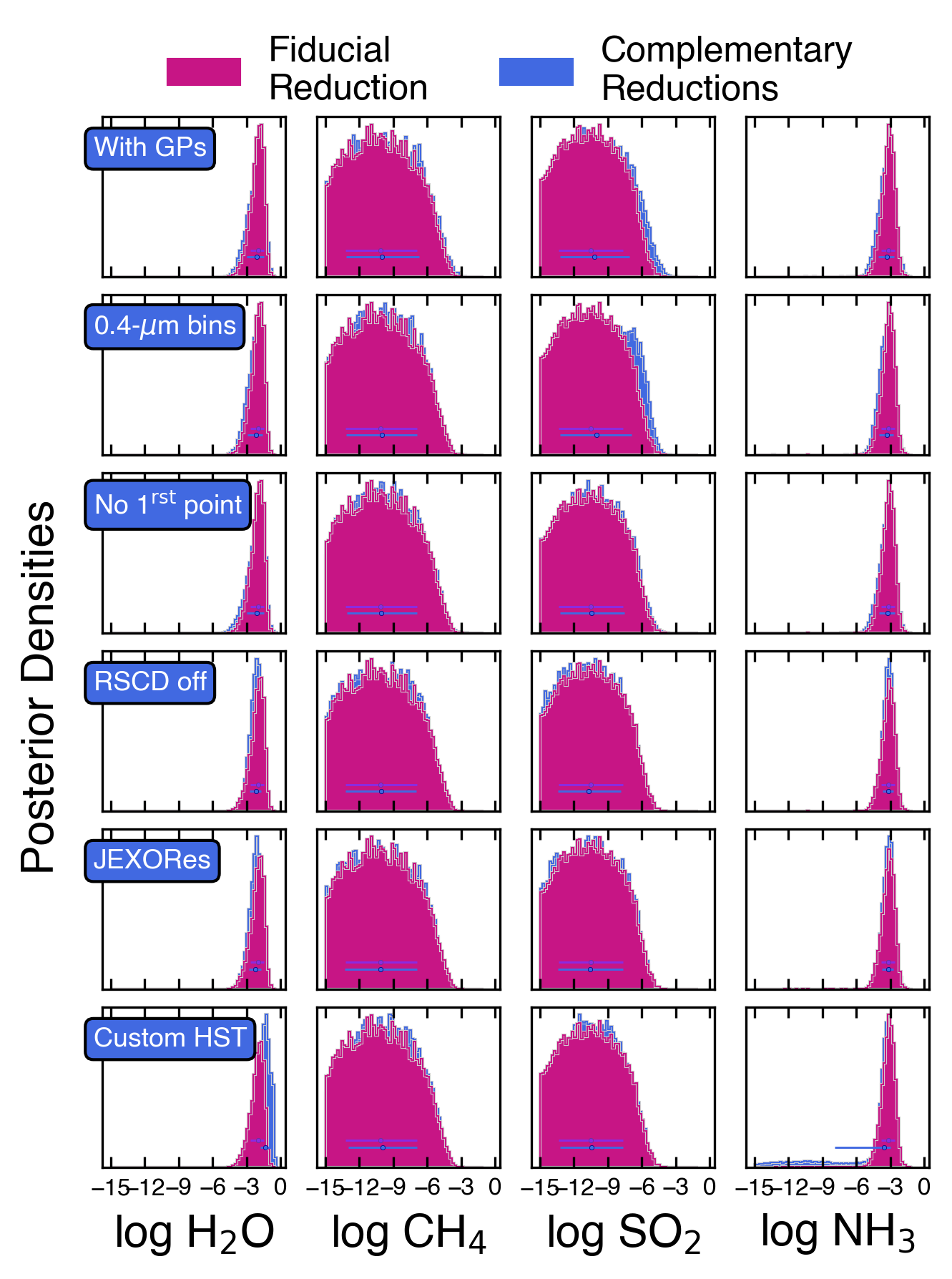}
    \caption{Posterior distributions of our retrieved abundances for H$_2$O, CH$_4$, SO$_2$ and NH$_3$ obtained for all 6 data reduction strategies. The fiducial reduction obtained with the \texttt{Eureka!} pipeline is depicted in purple, while results from other reduction strategies or pipelines are depicted in blue.}
    \label{fig:mol_constraints}
\end{figure*} 

\section{Atmospheric modeling}\label{app:atm_models}

\begin{figure*}
    \centering
    \includegraphics[width=0.4\linewidth]{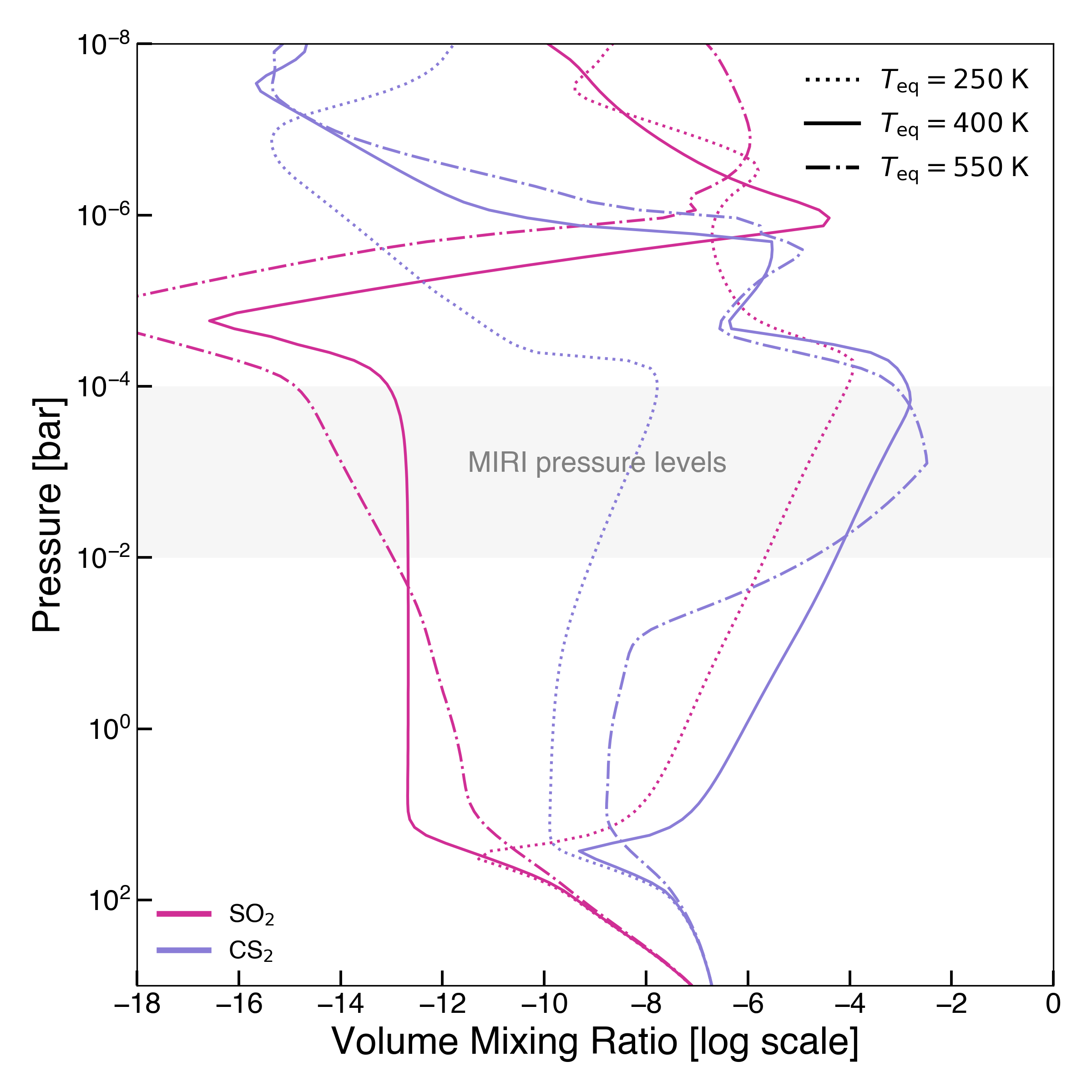}
    \includegraphics[width=0.4\linewidth]{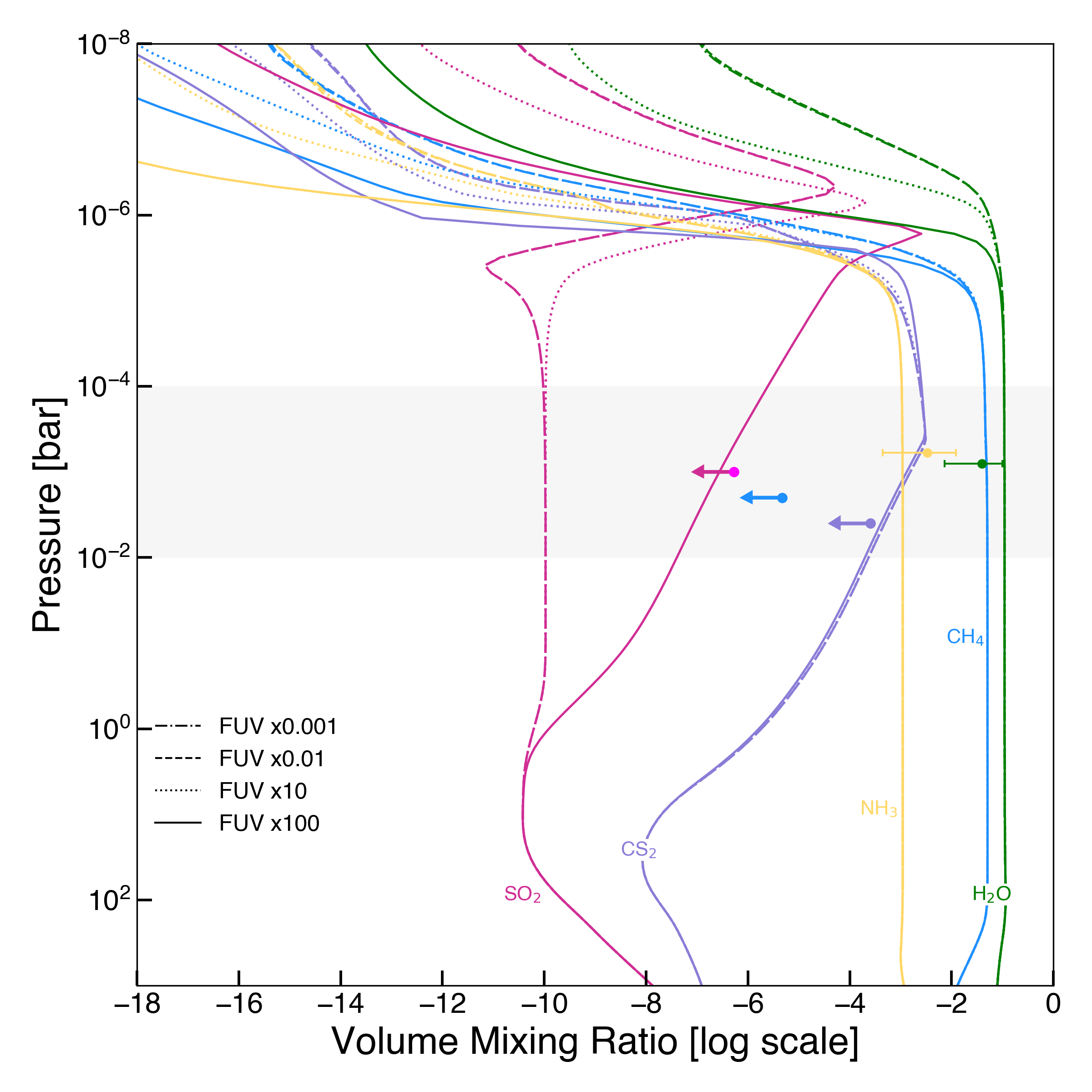}
    \caption{Vertical abundance profiles varying the equilibrium temperature and the FUV flux received by the planet. \textit{Left:} Vertical abundance profiles of SO$_2$ and CS$_2$ for three distinct low equilibirium temperatures. The pressures sounded by our JWST/MIRI transmission spectrum are shown by the grey area. \textit{Right:} Vertical abundance profiles for four different FUV fluxes received by the planet, scaled from the FUV flux of HD~106315.}
    \label{fig:models_low_teq}
\end{figure*}

As indicated in Sect.~\ref{sec:discussion}, we performed a series of atmospheric disequilibrium models of HD~106315~c using the \texttt{VULCAN} code \citep{tsai_vulcan_2017, tsai_comparative_2021}. One of the series of models consisted of varying the atmospheric metallicity and the equilibrium temperature to create an abundance map of SO$_2$ similarly to the work conducted by \cite{crossfield_mapping_2025}. We also provided an abundance map of CS$_2$ as both molecules are complementary tracers of temperature and metallicity. Our results show an enhanced abundance of SO$_2$ at very low temperatures ($\sim250-400$ K), where we would expect CS$_2$ to be the prominent sulphur-bearing species. To furthur investigate this, we computed the vertical abundance profiles of CS$_2$ and SO$_2$ for three distinct equilibrium temperatures (250, 400 and 500 K) that are shown in Fig.~\ref{fig:models_low_teq} and interpreted in Sect.~\ref{sec:shoreline}.

We additionally investigated the sensitivity of the molecular abundances to the incident far-ultraviolet (FUV) flux by varying the stellar irradiation relative to that received by HD~106315~c. Although changes in the molecular abundances become apparent in the upper atmosphere at high altitudes, they have little impact at the pressures probed with MIRI. Reducing the FUV flux by a factor of 1000 produces essentially the same abundance profiles as the nominal case, while increasing it by up to a factor of 10 likewise results in negligible changes over the MIRI-relevant pressure range. A more pronounced response is seen only for an FUV flux enhanced by a factor of 100, for which the SO$_2$ abundance is substantially modified and differences extend into the pressure range probed by MIRI. Thus, within the range of plausible FUV irradiation considered here, the predicted abundances in the observable atmosphere are largely insensitive to the incident FUV flux, with SO$_2$ becoming sensitive only under substantially enhanced irradiation.

\section{The stellar SED}\label{app:stellar_sed}

Here we provide some additional details regarding the stellar SED and its component input spectra.

\subsection{Optical and near-IR photometric SED}

We used the ARIADNE \citep{Vines2022} stellar SED fitting tool to analyse the photospheric emission at UV, optical and IR wavelengths. ARIADNE accesses publicly available photometric catalogs, and the GAIA catalog for the distance. For our analysis, we used the TYCHO, GAIA, Johnson V, TESS, 2MASS and WISE photometric points. These photometric points were fitted to a grid of PHOENIX \citep{Husser2013} model atmospheres. A best fit was found using a model with parameters T$_{\rm eff}$ = 6470~K (+130, -90), log~g~=~3.66 (+1.64,-0.16), and metallicity (dex)~=~-0.13 (+0.27 -0.35).  

\subsection{XMM-Newton X-ray spectrum}

To validate the spectral fit of the XMM-Newton data, we performed an independent test. We inferred the coronal temperature, relying on the correlation between the average coronal temperature $\bar{T}$ and the average stellar surface X-ray flux $F_{\rm X}$ \citep{johnstone2015} 
\begin{equation}\label{TFX}
    \bar{T} = 0.11\ F_{\rm X}^{0.26}.
\end{equation}
where $\bar{T}$ is in MK and $F_{\rm X}$ is in erg~cm$^{-2}$~s$^{-1}$. We use a stellar radius of $1.286R_{\odot} = 9 \times 10^{10}$~cm \citep{rodriguez2017}, and the X-ray luminosity derived in Sect.~\ref{sec:stellar_xray} to compute $F_{\rm X} = 2.417\times 10^4$~erg~cm$^{-2}$~s$^{-1}$. Our test shows that the expected coronal temperature is $\bar{T} = 1.5$~MK. Our fitted temperature is somewhat higher, although the discrepancy is moderate, and likely a result of enforcing an isothermal fit while stellar coronae normally show plasma distributed over a wide range of temperatures \citep{telleschi2005}.  

\subsection{Construction \& content of the SED}

\begin{figure*}
	\includegraphics[width=\textwidth]{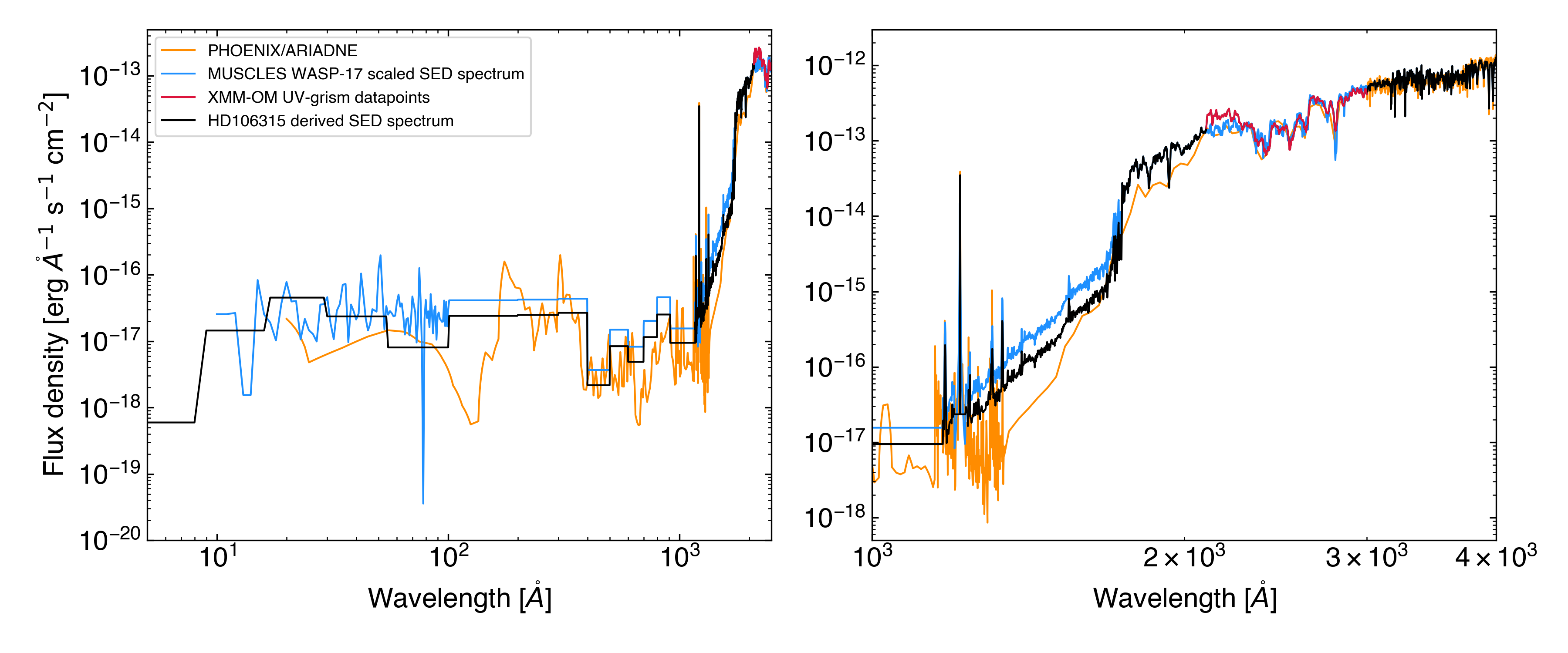}
    \caption{Spectral Energy Distribution of the host star HD~106315, at Earth distance. \textit{Left:} SED in the XUV range ($\sim 1$--$ 1000$ \AA. {\it Right:} SED in the XUV, FUV and NUV ranges ($\sim 1$--$4000$ \AA).}
    \label{fig:stellar_sed} 
\end{figure*}

The constructed SED covers the wavelength range from $\sim 5$ to $\sim 50000$ \AA\ (Fig.~\ref{fig:stellar_sed}), though the main relevance to the current study is the XUV+FUV+NUV range from $\sim 5$ to $\sim 4000$ \AA\ (Fig.~\ref{fig:xmm_nuv_grism}).
Table~\ref{tab:stellar_sed_inputs} lists the component input spectra, their wavelength ranges, origins, and other relevant details.
In selecting the origin of the input spectra, we adopt a similar priority scheme to MUSCLES \citep{france_muscles_2016}, using observational data of the target star if available, otherwise, in order: proxy spectra of similar stars, appropriately scaled; calculation, based on established correlations; stellar atmosphere models.
The inputs vary in terms of wavelength binning relative to the 1~\AA\ required for the final SED table; where needed, we have binned-up or interpolated the input fluxes.

We note a few additional points regarding the construction of the stellar SED. {\it (i)} The stellar Ly$\alpha$ flux was estimated using the Ly$\alpha$--X-ray correlation of \cite{johnstone2021} (their Eq.~24). For comparison, the correlation presented by \cite{linsky2020} (their Table~2, first equation) yields a Ly$\alpha$ flux approximately 40\% lower than the value derived from \cite{johnstone2021}.
{\it (ii)} The EUV spectrum was reconstructed using the wavelength-dependent Ly$\alpha$--EUV relations of \cite{linsky2014}, scaled to match the X-ray--EUV correlation of \cite{johnstone2021} over the 100--360~\AA\ wavelength range. Alternative empirical prescriptions available in the literature \citep[e.g.][]{linsky2014, king2018, sanz-forcada2025} can lead to integrated EUV flux estimates that differ by up to a factor of two. 
{\it (iii)} The FUV spectrum was represented using the MUSCLES SED of WASP-17 \citep{Behr2023}, scaled according to the ratio of the stellar bolometric fluxes. The MUSCLES SED itself incorporates a proxy FUV spectrum derived from observations of Procyon obtained with the SISTINE-II sounding-rocket spectrograph \citep{Behr2023}. We have applied an additional scaling factor of 0.5, derived by convolving the scaled WASP-17 SED with the GALEX FUV filter response and comparing the resulting synthetic flux with the measured GALEX FUV flux of HD~106315 \citep{loyd2020}\footnote{This factor is also comparable with the ratio of XMM-derived X-ray luminosities for HD~106315 and Procyon.}.

Apart from the flux scaling applied to the WASP-17 proxy spectrum, no additional renormalisation was performed to enforce continuity between adjacent spectral regions. The largest discontinuities occur at the EUV--FUV and FUV--NUV interfaces, where the flux levels differ by approximately a factor of two, although in opposite directions. The final SED is available as a FITS-format file\footnote{
SED FITS file and associated documentation: \texttt{<URL>}
}. The primary data columns contain the wavelength (\AA), flux density ($\mathrm{erg\ cm^{-2}\ s^{-1}\ \AA^{-1}}$), and associated uncertainty\footnote{
Flux-density uncertainties are provided where available and should be interpreted as indicative estimates of the signal-to-noise level rather than formal confidence intervals.
}. An additional column identifies the origin of each spectral segment (column ``INST.'' in Table~\ref{tab:stellar_sed_inputs}). The spectrum is sampled on a uniform wavelength grid with a bin width of 1~\AA.

\begin{sidewaystable*}
\renewcommand{\arraystretch}{1.25} 
\centering
\caption{Input components of the stellar SED.  }
\tiny
\begin{tabular}{lrrrrrlll}
\hline \hline
Region &
\multicolumn{2}{c}{Wavelength (\AA)} &
INST. &
Norm. &
Type &
Origin \& Notes &
Obs.\ Prog.\ ID &
  \\
  & start & end & \#  & factor  &  &  &     \\
\hline
  X-ray   & 5 & 100 & 1 &  & obs & [a] XMM; Sect.~\ref{sec:stellar_xray} + WebPIMMS & 0940590201 &   \\
  EUV   & 101 & 1170 & 2 & & calc & X-ray; Linsky+14, Johnstone+21 & n/a &   \\
  FUV   & 1171 & 1741 & 4 & 0.5 $\times$ & obs & GALEX phot. (Loyd+20)  & n/a &    \\
         &      &       &     & 11.69 & +proxy & + [c] MUSCLES WASP-17/Procyon as proxy (Behr+23)  & &   \\
   ... Ly-alpha & 1200 & 1230 & 8 &  & calc & X-ray; Linsky+14, Johnstone+21 & n/a &   \\
  NUV   & 1742 & 3171 & 16 &          & obs & [c] WASP-17 MUSCLES (HST-STIS-G230L; Behr+23)   & XMM 0940590201; Swift 16660; &   \\
                  &      &      &     & 11.69 & +proxy &  proxy scaled by ratio of $\rm F_{bol}$ & HST WASP-17 16166 &  \\
                  &      &      &     &       &        & + [a,d] XMM NUV spectra + Swift \& XMM photometry  &  &  \\
   ... NUV & 2100 & 3000 & 32 &       & obs    & [a,d] XMM-OM UV-grism  & XMM 0940590201; &   \\                  
~~~XMM-OM UV-grism &      &      &    &       &       & + XMM-OM+Swift-UVOT phot.  & Swift 16660  &  \\
  NUV-Visible     & 3172 & 5699 & 64 & 11.69 & proxy & [c] WASP-17 MUSCLES (HST-STIS-G430L; Behr+23)  & HST WASP-17 16166 &   \\
                  &      &      &    &       &       & proxy scaled by ratio of $\rm F_{bol}$  &  &   \\
  Visible-IR      & 5700 & 50000 & 128 & 1.0  & model & [e] PHOENIX/ARIADNE photospheric model tool  & n/a &   \\
         &      &       &     &  &       & (P.Daamen BSc Thesis; Vines \& Jenkins 2022) & &    \\
\hline
\end{tabular}
\tablefoot{
{\bf [a]} XMM-Newton Science Archive, ESA \url{https://nxsa.esac.esa.int/nxsa-web/\#search};
{\bf [b]} MAST/STScI HST Legacy Archive, HLSPs \url{https://hla.stsci.edu/hlaview.html};
{\bf [c]} MUSCLES website \& Archive at MAST \url{https://archive.stsci.edu/prepds/muscles};
{\bf [d]} GSFC HEASARC Swift Archive \url{https://heasarc.gsfc.nasa.gov/cgi-bin/W3Browse/swift.pl};
{\bf [e]} ARIADNE website \& code download \url{https://github.com/jvines/astroARIADNE/blob/master/README\_md}, performs fits of the model spectra to selected photometric bands, in this case 12, ranging from B to NIR.
}

\label{tab:stellar_sed_inputs}
\end{sidewaystable*}

\end{appendix}

\end{document}